\documentclass[fleqn,usenatbib]{mnras}

\usepackage{newtxtext,newtxmath}
\usepackage[T1]{fontenc}

\DeclareRobustCommand{\VAN}[3]{##2}
\let\VANthebibliography\thebibliography
\def\thebibliography{\DeclareRobustCommand{\VAN}[3]{##3}\VANthebibliography}

\usepackage{graphicx}	% Including figure files
\usepackage{amsmath}	% Advanced maths commands

\usepackage{caption}
\usepackage{subcaption}

\usepackage{booktabs}
\usepackage{threeparttable}

\title[REBELS-IFU: Metal-rich EoR galaxies]{REBELS-IFU: Evidence for metal-rich massive galaxies at $z\sim6-8$}

\author[Lucie E. Rowland]{
Lucie E. Rowland$^{1}$\thanks{E-mail:  lrowland@strw.leidenuniv.nl},
Mauro Stefanon$^{2,3}$,
Rychard Bouwens$^{1}$,
Jacqueline Hodge$^{1}$,
Hiddo Algera$^{4,5,6}$,
\newauthor Rebecca Fisher $^{7}$,
Pratika Dayal$^{8}$,
Andrea Pallottini$^{9}$,
Daniel P. Stark$^{10}$,
Kasper E. Heintz$^{11,12,13}$,
\newauthor Manuel Aravena$^{14,15}$,
Rebecca Bowler$^{7}$,
Karin Cescon$^{1}$,
Ryan Endsley$^{16}$,
Andrea Ferrara$^{9}$,
\newauthor Valentino Gonzalez$^{17,18}$,
Luca Graziani$^{19,20,21}$,
Cindy Gulis$^{22}$,
Thomas Herard-Demanche$^{1}$,
Hanae Inami$^{5}$,
\newauthor Andrès Laza-Ramos$^{2,3}$,
Ivana van Leeuwen$^{1}$,
Ilse de Looze$^{23}$,
Themiya Nanayakkara$^{24}$,
Pascal Oesch$^{13,11}$,
\newauthor Katherine Ormerod$^{7,25}$,
Nina S. Sartorio$^{23}$,
Sander Schouws$^{1}$,
Renske Smit$^{25}$,
Laura Sommovigo$^{26}$,
\newauthor Sune Toft$^{11,12}$,
John R. Weaver$^{27}$,
 Paul van der Werf$^{1}$
\\
$^{1}$ Leiden Observatory, Leiden University, P.O. Box 9513, 2300 RA Leiden,
The Netherlands\\
$^{2}$ Departament d’Astronomia i Astrofísica, Universitat de València, C. Dr. Moliner 50, E-46100 Burjassot, València, Spain\\
$^{3}$ Unidad Asociada CSIC “Grupo de Astrofísica Extragaláctica y Cosmología” (Instituto de Física de Cantabria - Universitat de València)\\
$^{4}$ Institute of Astronomy and Astrophysics, Academia Sinica, 11F of Astronomy-Mathematics Building, No.1, Sec. 4, Roosevelt Rd, Taipei 106216, Taiwan, R.O.C. \\
$^{5}$Hiroshima Astrophysical Science Center, Hiroshima University, 1-3-1 Kagamiyama, Higashi-Hiroshima, Hiroshima 739-8526, Japan \\
$^{6}$ National Astronomical Observatory of Japan, 2-21-1, Osawa, Mitaka, Tokyo, Japan \\
$^{7}$ Jodrell Bank Centre for Astrophysics, Department of Physics and Astronomy, School of Natural Sciences, The University of Manchester, Manchester M13 9PL, UK\\
$^{8}$ Kapteyn Astronomical Institute, University of Groningen, 9700 AV Groningen, The Netherlands\\
$^{9}$  Scuola Normale Superiore, Piazza dei Cavalieri 7, 56126, Pisa, Italy\\
$^{10}$ Steward Observatory, University of Arizona, 933 N Cherry Ave, Tucson, AZ 85721, USA\\
$^{11}$ Cosmic Dawn Center (DAWN), Copenhagen, Denmark\\
$^{12}$ Niels Bohr Institute, University of Copenhagen, Jagtvej 128, 2200
Copenhagen N, Denmark\\
$^{13}$ Department of Astronomy, University of Geneva, Chemin Pegasi 51, 1290 Versoix, Switzerland\\
$^{14}$Instituto de Estudios Astrof\'{\i}cos, Facultad de Ingenier\'{\i}a y Ciencias, Universidad Diego Portales, Av. Ej\'ercito 441, Santiago, Chile \\
$^{15}$ Millenium Nucleus for Galaxies (MINGAL)
$^{16}$Department of Astronomy, University of Texas, Austin, TX 78712, USA\\
$^{17}$ Departmento de Astronomia, Universidad de Chile, Casilla 36-D, Santiago 7591245, Chile\\
$^{18}$ Centro de Astrofisica y Tecnologias Afines (CATA), Camino del Observatorio 1515, Las Condes, Santiago 7591245, Chile\\
$^{19}$Dipartimento di Fisica, Sapienza, Università di Roma, Piazzale Aldo Moro 5, 00185 Roma, Italy \\
$^{20}$INFN, Sezione di Roma I, Piazzale Aldo Moro 2, 00185 Roma, Italy \\
$^{21}$INAF/Osservatorio Astronomico di Roma, Via di Frascati 33, 00078 Monte Porzio Catone, Italy \\
$^{22}$ Department of Astronomy, the Pennsylvania State University, 525 Davey Lab, University Park, PA 16802 \\
$^{23}$ Sterrenkundig Observatorium, Ghent University, Krijgslaan 281-S9, B-9000 Ghent, Belgium\\
$^{24}$ Centre for Astrophysics and Supercomputing, Swinburne University of Technology, P.O. Box 218, Hawthorn, 3122, VIC, Australia\\
$^{25}$ Astrophysics Research Institute, Liverpool John Moores University, 146 Brownlow Hill, Liverpool L3 5RF, United Kingdom\\
$^{26}$ Center for Computational Astrophysics, Flatiron Institute, 162 5th Avenue, New York, NY 10010, USA\\
Department of Astronomy, University of Massachusetts, Amherst, MA 01003, USA}

\date{Accepted XXX. Received YYY; in original form ZZZ}

\pubyear{\the\year{}}

\begin{document}
\label{firstpage}
\pagerange{\pageref{firstpage}--\pageref{lastpage}}
\maketitle

\begin{abstract}
Metallicity is a crucial tracer of galaxy evolution, providing insights into gas accretion, star formation, and feedback. At high redshift, these processes reveal how early galaxies assembled and enriched their interstellar medium. In this work, we present rest-frame optical spectroscopy of 12 massive ($\log(M_*/\mathrm{M_{\odot}})>9$) galaxies at $z\sim 6$-$8$ from the REBELS ALMA large program, observed with \textit{JWST} NIRSpec/IFU in the prism mode. These observations span emission lines from [O \textsc{ii}]$\lambda$3727,9 to [S \textsc{ii}]$\lambda$6716,31, providing key information on nebular dust attenuation, ionisation states, and chemical abundances. We find lower O32 ratios (average $\sim3.7$) and [OIII]$\lambda$5007 equivalent widths (average ${EW_{[OIII]}}\sim390$\AA) than are generally found in existing large spectroscopic surveys at $z>6$, indicating less extreme ionising conditions. Strong-line diagnostics suggest that these systems are some of the most metal-rich galaxies observed at $z>6$ (average $Z_{\mathrm{gas}}\sim 0.4 Z_{\odot}$), including sources with near-solar oxygen abundances, in line with their high stellar masses (average $\log{M_*/\mathrm{M_{\odot}}}\sim9.5$). Supplementing with literature sources at lower masses, we investigate the mass-metallicity and fundamental metallicity relations (MZR and FMR, respectively) over a 4 dex stellar mass range at $6<z<8$. In contrast to recent studies of lower-mass galaxies, we find no evidence for negative offsets to the $z=0$ FMR for the REBELS galaxies. This work demonstrates the existence of chemically-enriched galaxies just $\sim1$ Gyr after the Big Bang, and indicates that the MZR is already in place at these early times, in agreement with other recent $z>3$ studies.

\end{abstract}


%%%%%%%%%%%%%%%%%%%%%%%%%%%%%%%%%%%%%%%%%%%%%%%%%%

%%%%%%%%%%%%%%%%% BODY OF PAPER %%%%%%%%%%%%%%%%%%

\section{Introduction}
\label{sec:introduction}
Understanding the formation and evolution of galaxies in the early Universe is a central goal of modern astrophysics. Over the past decade, remarkable progress has been made in extending the redshift frontier, with the highest-redshift galaxy observations advancing from $z=8.68$, achieved by a Ly$\alpha$ detection with Keck/MOSFIRE in 2015 (\citealt{zitrin_lyman_2015,roberts-borsani_z_2016}), to $z=14.18$ in 2024 with \textit{JWST} and ALMA (\citealt{schouws_detection_2024,carniani_spectroscopic_2024}) – just $\sim$300 million years after the Big Bang. With \textit{JWST} significantly increasing the number of observed galaxies at $z\gtrsim10$ (e.g. \citealt{curti_jades_2023, robertson_identification_2023, bunker_jades_2023, carniani_spectroscopic_2024}), studies have uncovered an unexpected population of luminous ($M_{\text{UV}} < -20$) galaxies already in place at these early cosmic times (e.g. \citealt{naidu_two_2022,castellano_early_2022,casey_cosmos-web_2024,zavala_luminous_2024}). Even before \textit{JWST}, there were notable discoveries of massive galaxies ($M_*\sim 10^9$--$10^{11} \mathrm{M_{\odot}}$, e.g. \citealt{strandet_ism_2017, marrone_galaxy_2018,banados_800-million-solar-mass_2018, bouwens_reionization_2022}) and substantial dust reservoirs (\citealt{watson_dusty_2015, tamura_detection_2019, schouws_significant_2022,inami_alma_2022}) at slightly later cosmic times ($z\sim6$-$8$). These observations provide valuable insights into the early assembly of galaxies, but many of these studies lack robust constraints on the metallicity of the systems.

Metallicity is a key diagnostic for understanding how galaxies have formed and evolved. Metals provide insights into a galaxy's star formation history, gas inflows and outflows, and the feedback processes that regulate its growth (e.g. \citealt{maiolino_re_2018}). Inflows bring in pristine gas from the surrounding intergalactic medium (IGM) that, while diluting metallicity in the short term, fuels future star formation. This gas eventually forms stars, where nuclear fusion transforms hydrogen and helium into heavier elements. Massive, newly-formed stars return energy and metals to the interstellar medium (ISM), and can drive powerful outflows that carry gas and metals out of the galaxy. These expelled metals can escape the galaxy’s gravitational pull to enrich the IGM or reaccrete to enrich the infalling gas. This dynamic cycling of baryons in and out of galaxies shapes their stellar mass, metallicity, and star formation rate (SFR) (e.g., \citealt{peeples_constraints_2011,lilly_gas_2013,torrey_evolution_2019}). Galaxy metallicities are therefore one of the fundamental observational quantities that provide information about galaxy evolution, but they were largely unconstrained at $z>3$ prior to the launch of \textit{JWST} (but note e.g. \citealt{jones2020}). 

The measurement of the gas-phase metallicities of star-forming galaxies, traced by their oxygen abundance ($12+\log(\mathrm{O/H})$), typically requires the detection of rest-frame optical emission lines that are absorbed by the Earth’s atmosphere at $z\gtrsim3$. \textit{JWST}, with its unprecedented high-angular resolution and sensitivity in the near- and mid-infrared, has now begun to address this gap, offering routine detections of rest-frame optical emission lines at higher redshifts (e.g., \citealt{arellano-cordova_first_2022,curti_chemical_2022,katz_first_2023,taylor_metallicities_2022,trump_physical_2022,schaerer_first_2022,brinchmann_high-z_2023,carnall_first_2023}). This has enabled key metallicity scaling relations, such as the mass-metallicity relation (MZR) and fundamental metallicity relation (FMR), to be constrained out to $z\sim10$ (e.g. \citealt{heintz_dilution_2023,nakajima_jwst_2023, curti_jades_2023, chemerynska_extreme_2024, sarkar_unveiling_2024,venturi_gas-phase_2024}). However, the observational surveys used in these latest studies typically poorly sample the high-mass ($M_*>10^9 \mathrm{M_{\odot}}$) end of these relations, mainly due to limitations of survey volumes and due to the rarity of such massive systems at $z>6$. Extending these metallicity studies to massive high-$z$ galaxies is particularly important, as these galaxies likely represent both the descendants of the surprisingly bright galaxies discovered at $z>10$ and the building blocks of the most massive structures observed at later times (e.g. \citealt{hashimoto_onset_2018, hygate_alma_2023, narayanan_ultraviolet_2024}). Studying the metal content of massive $z>6$ galaxies therefore provides a valuable foundation for exploring how early galaxies could have built up such high stellar and dust masses so soon after the Big Bang.

To address these challenges, a sample of spectroscopically-confirmed massive galaxies in the Epoch of Reionisation (EoR) is necessary. Amongst the largest contributions to this effort is the REBELS (Reionisation Era Bright Emission Line Survey,  \citealt{bouwens_reionization_2022}) ALMA large program, which has significantly expanded the sample of EoR galaxies with robust dust continuum and [C \textsc{ii}] 158 $\mu$m detections. This survey targets 40 UV-bright (\( M_\mathrm{UV} < -21.5 \)) galaxies at \( z \sim 6.5 - 9 \), resulting in a sample of 16 galaxies with confirmed dust continuum detections and 25 galaxies with [C \textsc{ii}] detections (Schouws et al. in prep, \citealt{inami_alma_2022}). From this survey, 12 of the [C \textsc{ii}]-brightest galaxies were targeted in \textit{JWST} Cycle 1 programs (GO 1626: PI M. Stefanon and GO 2659: PI J. Weaver) with the NIRSpec/IFU. In this work, we present the emission line fitting of these NIRSpec data, which we use to derive the oxygen abundances and other key ionised gas properties of these 12 galaxies.

This paper is structured as follows: In Section \ref{sec:data}, we describe the data products from the REBELS program and the \textit{JWST} NIRSpec observations. Section \ref{sec:emission line fitting} details the methodology for extracting integrated spectra and fitting emission lines. In Section \ref{sec:ionised gas properties}, we detail the derivation of the ionised gas properties of the sample, and in Section \ref{sec:sed fitting} we summarise the SED fitting used to derive the stellar population properties (with more details given in a subsequent paper, Stefanon et al. in prep). We discuss the implications of our findings for the ionised gas properties of the REBELS galaxies, and put them into context by studying the MZR and FMR at $z\sim6$-$8$ in Section \ref{sec:discussions}. Finally, we summarise our work and conclusions in Section \ref{sec:conclusions}. Throughout this work, we assume a standard $\Lambda$CDM cosmology, with $H_0=70$, km s$^{-1}$ Mpc$^{-1}$, $\Omega_m=0.30$ and $\Omega_\Lambda=0.70$. We further adopt a \cite{kroupa_variation_2001} initial mass function, and take solar abundance to be $12+\log(\mathrm{O/H})=8.69$ (following \citealt{asplund_chemical_2009}).

\section{Data}
\label{sec:data}
\begin{table}
\centering
\def\arraystretch{1.25}
\addtolength{\tabcolsep}{-2.5pt}
\caption{Summary of the properties derived from pre-existing ground-based observations of our sample.}
\begin{tabular}{lcccccccc}
\hline
Galaxy & RA & Dec & $z_{\mathrm{[C \textsc{ii}]}}$ & $L_{\mathrm{[C \textsc{ii}]}}$ & $L_{\mathrm{IR}}$  \\ 
 & & &  & $10^{8}\mathrm{L_{\odot}}$ & $10^{11}\mathrm{L_{\odot}}$ \\
\hline
REBELS-05 &  02:18:11.51 & -05:00:59.3 & 6.496 & $6.9^{+0.8}_{-0.9}$ & $3.2^{+1.9}_{-1.2}$ \\
REBELS-08 & 02:19:35.13 & -05:23:19.2 & 6.749 & $7.4^{+1.0}_{-1.1}$ & $5.2^{+3.0}_{-2.0}$\\
REBELS-12 &  02:25:07.94 & -05:06:40.7 & 7.349 & $10\pm4$ & $2.8^{+29}_{-1.4}$$^{\mathrm{b}}$\\ 
REBELS-14 &  02:26:46.19 & -04:59:53.5 & 7.084 & $3.7^{+1.1}_{-1.0}$ & $3.3^{+2.0}_{-1.4}$\\
REBELS-15 & 02:27:13.11 & -04:17:59.2 & 6.875 & $1.9^{+0.5}_{-0.4}$ & $<3.6$\\
REBELS-18 &  09:57:47.90 & 02:20:43.7 & 7.675 & $11^{+1.0}_{-0.9}$ & $3.5^{+2.0}_{-1.3}$\\ 
REBELS-25 &  10:00:32.32 &  01:44:31.3 & 7.306 & $17\pm2$$^\mathrm{a}$ & $5.0_{-1.0}^{+2.9}$$^{\mathrm{c}}$\\
REBELS-29 & 10:01:36.85 &  02:37:49.1 & 6.685 & $5.5^{+1.0}_{-0.9}$ & $2.9^{+1.7}_{-1.1}$\\
REBELS-32 & 10:01:59.07 &  01:53:27.5 & 6.729 & $7.9^{+0.8}_{-0.9}$ & $3.1^{+1.9}_{-1.3}$\\
REBELS-34 & 10:02:06.47 &  02:13:24.2 & 6.633 & $6.9^{+2.6}_{-1.9}$ & $<3.8$\\
REBELS-38 & 10:02:54.05 &  02:42:12.0 & 6.577 & $17^{+1.6}_{-1.5}$ & $3.0^{+1.7}_{-0.6}$$^{\mathrm{b}}$\\
REBELS-39 & 10:03:05.25 & 02:18:42.7 & 6.847 & $7.9^{+2.5}_{-2.4}$ & $4.2^{+2.4}_{-1.6}$\\
\hline
\end{tabular}
%\begin{tablenotes}
\textbf{Notes:} Right Ascension (RA), Declination (Dec) and spectroscopic redshifts ($z_{\mathrm{[C \textsc{ii}]}}$) in columns 1, 2, and 3, respectively, are taken from \cite{bouwens_reionization_2022}. Col. 5 lists the [C \textsc{ii}]158$\mu$m luminosities from S. Schouws et al. (in prep), and Col. 6 lists the IR IR luminosities taken from \cite{inami_alma_2022} for detections in ALMA's band 6 assuming an emissivity index $\beta=2.0$.$^a$ The [C \textsc{ii}] luminosity for REBELS-25 is taken from \cite{hygate_alma_2023}. $^b$ The IR luminosities for REBELS-12 and REBELS-38, which have continuum detections/limits in two bands with ALMA, are taken from the fiducial case in \cite{algera_cold_2024} ($\beta=2.0$, optically thin). $^c$ The IR luminosity of REBELS-25 is taken from from \cite{algera_accurate_2024}.
%\end{tablenotes}
\label{tab:REBELS properties}
\end{table}

The 12 galaxies analysed in this work were selected from the REBELS ALMA large program (\citealt{bouwens_reionization_2022}, Schouws et al in prep) and were observed with \textit{JWST} NIRSpec/IFU in Cycle 1 (GO-1626, PI M. Stefanon), with one source (REBELS-18) observed in GO 2659 PI J. Weaver. All 12 of these galaxies have [C \textsc{ii}] detections at $>7\sigma$ from the REBELS large program, and all but two have continuum detections at rest-frame $\sim 160 \mu$m (\citealt{inami_alma_2022}). We summarise key properties from the REBELS ALMA large program and pre-existing photometry in Table \ref{tab:REBELS properties}.  The \textit{JWST} observations of these sources were performed using the prism mode, covering an observed wavelength range of 0.6–5.3 $\mu$m with a spectral resolution of $R \sim 100$, for 1700 seconds of exposure per source (for REBELS-18, the exposure time is $\sim1.7$ hours). A full description of the reduction of the JWST data will be provided in Stefanon et al. (in prep). Briefly, we employed a customized version of the JWST pipeline, incorporating the \texttt{GRIZLI} (\citealt{brammer_grizli_2023}) implementation of Stage 1 with additional steps for cosmic ray masking using \texttt{ASTROSCRAPPY} (\citealt{van_dokkum_cosmic-ray_2001}) and manual masking of hot edge pixels. For Stages 2 and 3, we used the default pipelines and parameters (albeit adopting 0.08 arcsec for the pixel scale). The final data cubes were background-subtracted by masking each main source and applying a 2D linear interpolation to reconstruct the background for each source.

\subsection{Mask selection}
\label{sec:mask selection}
\begin{figure*}
    \centering
    \includegraphics[width=0.9\textwidth]{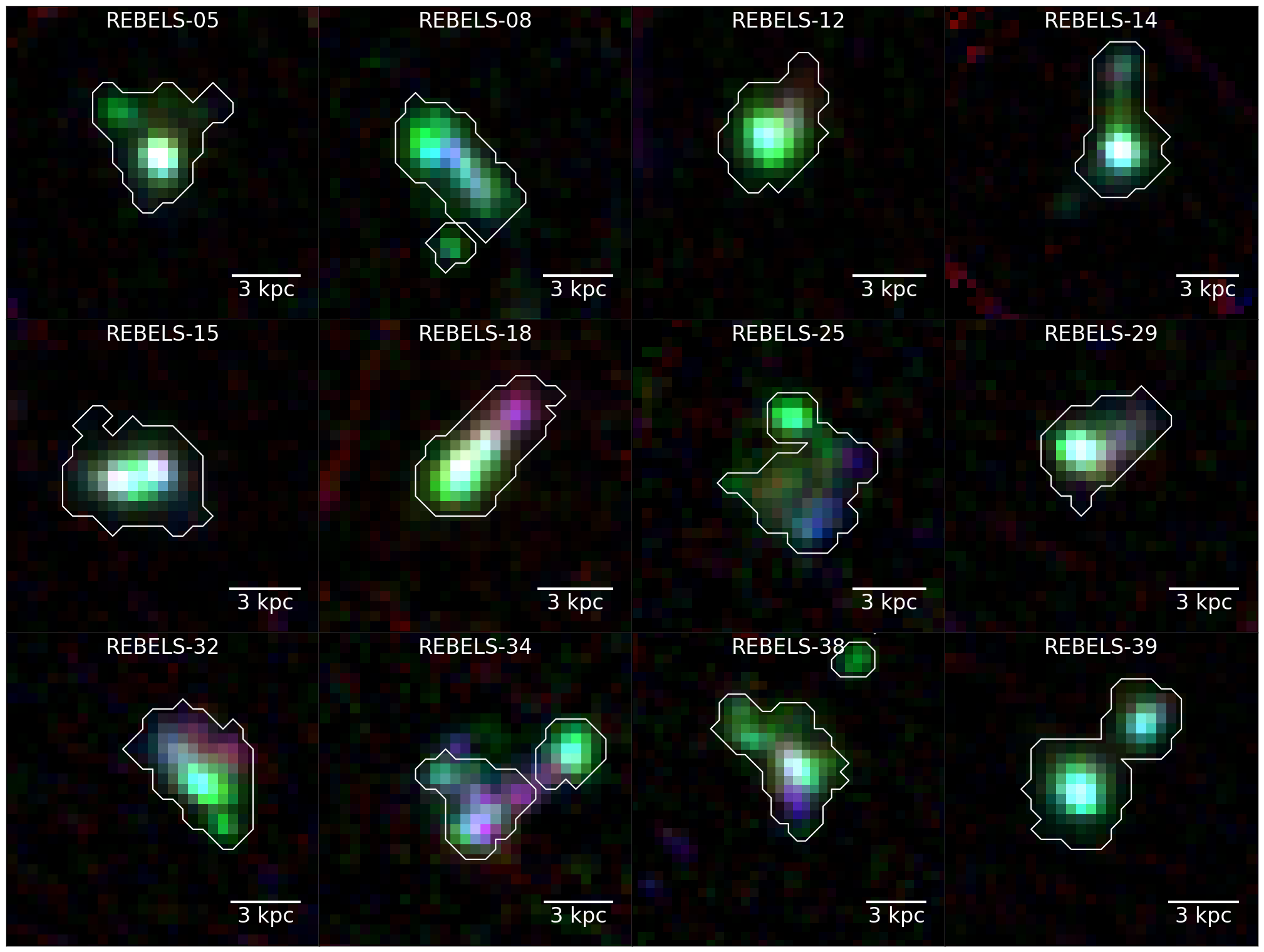}
    \caption{Three-colour composite images of each of the 12 REBELS galaxies targeted in the Cycle 1 \textit{JWST} program analysed in this work. Red corresponds to rest-frame optical ($0.37<\lambda_{\mathrm{rest}}< 0.855 \mu$m, excluding emission lines), green to [O \textsc{iii}]$\lambda$5007 and blue to rest-frame UV emission ($0.12<\lambda_{\mathrm{rest}}<0.37 \mu$m, excluding emission lines). The white contours are used to extract the integrated spectrum of each galaxy, as described in the text.}
    \label{fig:rgb plots}
\end{figure*}

In this work, we focus on the integrated spectrum and global properties of each galaxy. For the extraction of 1D spectra, we create source-specific apertures by combining masks that capture flux emission at $>3\sigma$ in various wavelength ranges, including the rest-frame UV, rest-frame optical, and key emission lines such as [O \textsc{ii}]$\lambda$3727, H$\beta$, [O \textsc{iii}]$\lambda$4959,5007, and H$\alpha$. These masks were merged to define the final aperture for each galaxy, covering all regions with significant flux detection. A full description of the masking procedure will appear in a forthcoming paper (Stefanon et al. in prep). In Figure \ref{fig:rgb plots}, we show three colour-composite images of the 12 galaxies with the masks used to extract each spectrum in white, and in Figure \ref{fig:all spectra} we plot the integrated spectra extracted from these masks. In many cases, the emission is clumpy and irregular. For the purpose of this work, we derive properties based on the integrated spectrum of all the pixels contained within these masks, with further spatially resolved analyses to follow in subsequent works. We therefore caution that some sources, assumed in this work to be single objects, may in fact be merging or interacting systems.

\section{Emission line measurements}
\label{sec:emission line fitting}

In order to model the observed emission lines, we first continuum-subtract the integrated spectra extracted from the masks. To do this, we fit a third-order polynomial to a portion of each spectrum (from $\lambda>0.3 \mu$m in the rest-frame), excluding wavelengths where emission lines are detected, using the \texttt{PYSPECKIT} Python package (\citealt{ginsburg_pyspeckit_2022}). {The continuum-line fits are show in Appendix \ref{sec:appendix emission line fitting}. We then fit all key nebular emission lines with $> 1\sigma$ emission at the expected wavelength centroid based on the redshift of each galaxy in the NIRSpec wavelength range using the Gaussian model within \texttt{PYSPECKIT}. To reduce the number of free parameters, we fix the centroid and width of each of the key emission lines according to the redshift and line spread function (LSF), since at the resolution of the prism grating the widths of the lines are dominated by the spectral resolution (see Appendix \ref{sec:appendix emission line fitting}). We also test subtracting the stellar continuum from the best-fit SED (discussed in Section \ref{sec:sed fitting}), in order to account for stellar absorption at the hydrogen recombination lines, but note that this makes a negligible change to the derived emission line fluxes, even for the Balmer lines, since we see very little stellar absorption. This suggests that the emission is dominated by bright, young stellar populations, which is perhaps not surprising considering these galaxies are UV-selected and amongst the most UV-bright in the EoR. %\textcolor{red}{Note: if it is preferred, I can easily use this stellar continuum subtraction method instead. The difference in emission line fluxes seems to be incredibly minor - is this surprising?}

Due to the coarse resolution of the prism (R $\sim100$), many of the key emission lines are blended. In addition, it is not possible to identify and fit broad components to the emission lines at the current spectral resolution. We therefore fit the following blended emission lines as single Gaussians: [O\textsc{ii}]$\lambda$3727+[O\textsc{ii}]$\lambda$3729, [Ne\textsc{iii}]$\lambda$3869+He\textsc{i}+H$\zeta$, H$\gamma$+[O\textsc{iii}]$\lambda$4363+[Fe\textsc{ii}]$\lambda4360$ and [S\textsc{ii}]$\lambda$6716+[S\textsc{ii}]$\lambda$6731. Note that we assume that the relative contribution of the fainter He \textsc{i} and H$\zeta$ lines is negligible, in comparison to the uncertainties, to the total flux around the [Ne \textsc{iii}] emission line, and hereafter we assume that this blended feature is equivalent to only the [Ne \textsc{iii}] flux. The H$\alpha$ emission line is also blended with the [N \textsc{ii}]$\lambda$6548,84 doublet at the resolution of the prism. However, for the 8 sources at $z\lesssim7$ where H$\alpha$ is detected, the line is visibly asymmetric. We therefore fit a triple-Gaussian to the blended lines in order to determine the Balmer decrement and obtain attenuation-corrected emission line fluxes. To reduce the number of free parameters in this triple Gaussian fit, as well as fixing the line widths and the line centres, we also tie the flux ratio of the [N \textsc{ii}] doublet to 3.049 (\citealt{storey_theoretical_2000}), so that the only free parameters are the amplitudes of H$\alpha$ and [N \textsc{ii}]$\lambda6584$. For the sources where [N \textsc{ii}]$\lambda6584$ is detected at  $\gtrsim3\sigma$, we find that this triple-Gaussian model produces a better fit than a single Gaussian by comparing the resulting Akaike Information Criterion (AIC) and/or reduced chi-squared values, justifying our use of this method. Since this blended emission means that the [N \textsc{ii}] and H$\alpha$ fluxes are correlated, we propagate these correlated uncertainties when determining the ratio of [N \textsc{ii}] to H$\alpha$ in subsequent sections. An example of this triple-Gaussian fit is plotted in Figure \ref{fig:deblending H-alpha}. We also fix the flux ratio of the [O \textsc{iii}]$\lambda$4959,5007 doublet to the theoretical value of 2.98, but note the detected lines are consistent with this theoretical value within the uncertainties if not fixed, reinforcing the robustness of our data reduction.

We add that the fits are performed twice with \texttt{PYSPECKIT}, so that the residuals in each spectrum are used to estimate the uncertainties in the final fitted parameters (see \citealt{ginsburg_pyspeckit_2022}). If the error spectrum from the `\texttt{ERR}' extension of the data cube is instead used, the uncertainties typically decrease by a factor of three. We use the larger uncertainties in the integrated fluxes to be conservative. A similar discrepancy with the reported flux uncertainties in the `\texttt{ERR}' extension is also noted in \cite{ubler_ga-nifs_2023}.

\begin{figure*}
    \centering
    \includegraphics[width=0.9\textwidth]{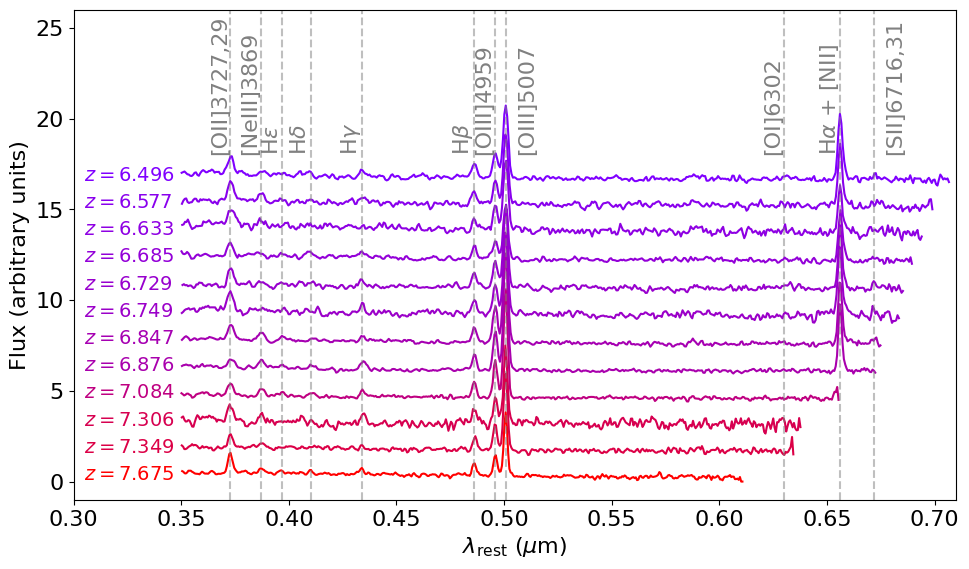}
    \caption{Integrated NIRSpec spectrum of all 12 REBELS targeted with our IFU program, extracted from a target-specific aperture mask shown in Figure \ref{fig:rgb plots}.  Each spectrum is only shown over the range $\lambda_{\mathrm{rest}}=0.35\mu$m to $\lambda_{\mathrm{obs}}=5.3\mu$m to highlight the key detected rest-frame optical emission lines, labelled with the grey dashed lines and text. The fluxes are normalised to the H$\beta$ flux of each galaxy, and offset along the vertical axis for clarity. Galaxies at the lower end of the redshift range of our REBELS targets are plotted in purple while galaxies at the higher end of the range are plotted in red.}
    
    \label{fig:all spectra}
\end{figure*}

\begin{figure}
    \centering
    \includegraphics[width=0.48\textwidth]{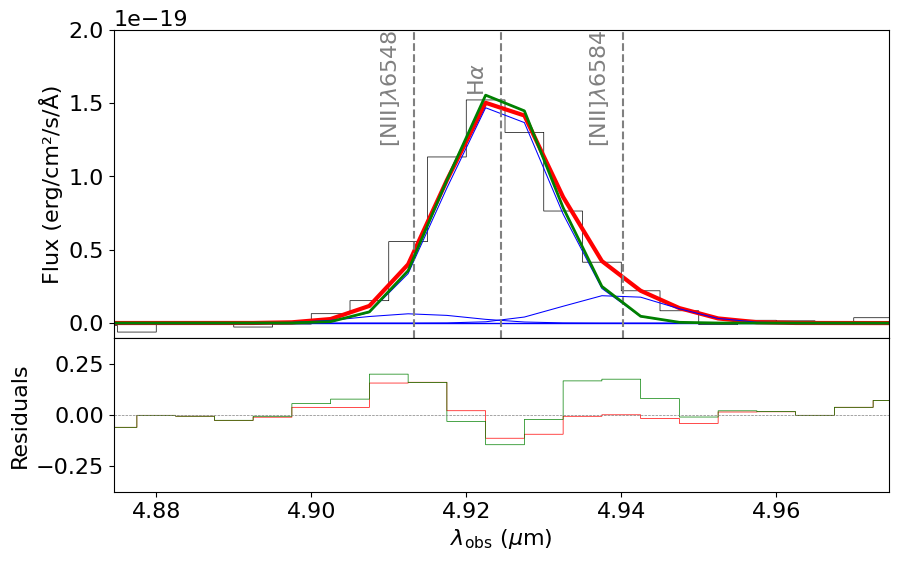}
    
    \caption{An example of the triple-Gaussian fit to the [N \textsc{ii}]$\lambda6548$, H$\alpha$, and [N \textsc{ii}]$\lambda6584$ blended spectral feature for REBELS-05 is shown in red, with each component plotted in blue, in the top panel. We also show a single Gaussian-fit to H$\alpha$ only in green. In the bottom-panel, we plot the residuals of the triple-Gaussian fit (red) and the single Gaussian-fit (green). } 
    \label{fig:deblending H-alpha}
\end{figure}

\subsection{Reddening correction}
\label{sec:reddening correction}

The NIRSpec prism observations for eight of our targets provide us with coverage of both the H$\alpha$ and H$\beta$ lines, allowing us to estimate reddening corrections on the basis of the Balmer decrement.  For each source, the nebular colour excess, $E(B-V)_{\mathrm{neb}}$, is calculated from the Balmer decrement of each galaxy assuming Case B recombination at $T \sim 10^4$ K and electron density $n_e \sim 100 ~\mathrm{cm}^{-3}$ with an intrinsic ratio of 2.86 for H$\alpha$/H$\beta$ (typical for HII regions, \citealt{hummer_recombination-line_1987, osterbrock_astrophysics_2006}). This ratio is weakly sensitive to density and temperature (e.g., varying from 2.86 to 2.81 for $n_e=10^2 - 10^6$ cm$^{-3}$ at $T=10^4$ K, with detailed values in, for example, \citealt{dopita_astrophysics_2003} and \citealt{osterbrock_astrophysics_2006}). These small variations are negligible compared to dust effects, making the Balmer decrement a reliable diagnostic of reddening. We then make use of the measured Balmer decrements and the \citet{calzetti_dust_2000} attenuation curve to correct the observed fluxes of various emission lines.

The continuum-subtracted, attenuation-corrected fluxes for these eight sources, which are listed in Table \ref{tab:corrected flux catalogue}, are then used for all following calculations with the uncertainties returned by \texttt{PYSPECKIT} propagated through. We also determine the nebular visual attenuation, $A_\mathrm{V,neb}$, for each galaxy from the product of the colour excess and the total-to-selective attenuation, for which we adopt the value of 4.1$\pm$0.8 derived by \citet{calzetti_dust_2000}. We note that the results presented in this paper do not significantly change if we had instead adopted a Milky Way attenuation curve (\citealt{cardelli_relationship_1989}) or LMC attenuation curve (\citealt{fitzpatrick_correcting_1999}), and a more detailed analysis of the dust attenuation within these galaxies will be the focus of a subsequent paper (Fisher et al. submitted). The main findings of this paper are also consistent when the attenuation curve empirically derived from the NIRSpec data for each source (Fisher et al. submitted) is used.

For the four galaxies where H$\alpha$ is beyond the wavelength coverage of NIRSpec, we do not attempt a reddening correction of the emission line fluxes based on the H$\gamma$/H$\beta$ ratio, since these lines are detected at lower SNR than H$\alpha$, increasing the uncertainty in the derived $A_{\mathrm{V, neb}}$, and because the observed H$\gamma$/H$\beta$ ratios are consistent with unphysical values (assuming Case B recombination) for the majority of the sample, likely as a consequence of the low SNR detections. More details are given in Appendix \ref{sec:appendix balmer decrements}. We therefore give the non-attenuation-corrected values for these four $z\gtrsim7$ galaxies in Table \ref{tab:non-corrected flux catalogue}.

\begin{table*}
\centering
\caption{Attenuation-corrected emission line fluxes for the eight galaxies where H$\alpha$ lies within the NIRSpec wavelength coverage. Fluxes are quoted in units of $10^{-18}$ erg s$^{-1}$ cm$^{-2}$ with their uncertainties. Where the SNR$<3$, we report the $3\sigma$ upper limits on the flux. Dashes represent emission lines that are redshifted out of the NIRSpec wavelength range.}
\begin{tabular}{lccccccccccc}
\hline
Galaxy & [O \textsc{ii}]$^{a}$ & [Ne \textsc{iii}]3869$^{b}$ & H$\epsilon$ & H$\delta$ & H$\gamma^{c}$ & H$\beta$ & [O \textsc{iii}]5007 & [O \textsc{i}]6302 & H$\alpha^{d}$ & [N \textsc{ii}]6584$^{d}$ & [S \textsc{ii}]$^{e}$ \\ 
\hline
REBELS-05 & 38 $\pm$ 4 & < 8.8 & < 6.7 & < 14 & 9.2 $\pm$ 2.5 & 19 $\pm$ 2 & 93 $\pm$ 2 & < 4.4 & 53 $\pm$ 2 & 6.8 $\pm$ 1.6 & < 5.4 \\ 
REBELS-08 & 44 $\pm$ 5 & < 13 & < 11 & < 8.2 & 13 $\pm$ 3 & 18 $\pm$ 3 & 100 $\pm$ 2 & < 4.8 & 51 $\pm$ 2 & 8.3 $\pm$ 1.6 & 5.3 $\pm$ 1.6 \\ 
REBELS-15 & 19 $\pm$ 4 & 16 $\pm$ 3 & < 6.2 & < 7.5 & 16 $\pm$ 3 & 24 $\pm$ 2 & 180 $\pm$ 2 & < 2 & 69 $\pm$ 2 & < 9.4 & - \\ 
REBELS-29 & 11 $\pm$ 1 & < 3.2 & < 3.2 & 3.8 $\pm$ 1.0 & 5.4 $\pm$ 1.5 & 8.1 $\pm$ 0.9 & 35 $\pm$ 1 & < 1.9 & 23 $\pm$ 1 & 3.3 $\pm$ 0.7 & < 2.5 \\ 
REBELS-32 & 23 $\pm$ 2 & < 5.9 & < 4.4 & < 5.4 & 6.5 $\pm$ 1.5 & 13 $\pm$ 1 & 71 $\pm$ 1 & < 1.9 & 37 $\pm$ 1 & 3 $\pm$ 1 & 3.2 $\pm$ 0.9 \\ 
REBELS-34 & 29 $\pm$ 4 & < 9.3 & < 8.2 & < 5.5 & < 4.6 & 8.8 $\pm$ 1.8 & 64 $\pm$ 2 & < 5.6 & 25 $\pm$ 1 & < 5.6 & < 17 \\ 
REBELS-38 & 59 $\pm$ 5 & 23 $\pm$ 4 & < 8.8 & < 8.2 & 14 $\pm$ 5 & 21 $\pm$ 3 & 110 $\pm$ 31 & < 6.4 & 60 $\pm$ 2 & 6.2 $\pm$ 1.9 & < 15 \\ 
REBELS-39 & 29 $\pm$ 4 & 13 $\pm$ 3 & < 7 & < 6.1 & < 9.6 & 20 $\pm$ 2 & 150 $\pm$ 2 & < 3 & 57 $\pm$ 2 & < 6.6 & < 4.2 \\ 
\hline
\end{tabular}
\begin{tablenotes}
\item $^a$ Blended [O \textsc{ii}]$\lambda$3727,9 doublet
\item $^b$ Blended with He \textsc{i} and H$\zeta$, however for the purpose of this work we assume that the contribution from these fainter lines is negligible in comparison to the uncertainties. 
\item $^c$ Blended with the auroral [O \textsc{iii}]$\lambda$4363 line and the [FeII]4360 line, however for the purpose of this work we assume that the contribution from these fainter lines is negligible in comparison to the uncertainties. 
\item $^d$ The H$\alpha$ and [N \textsc{ii}]$\lambda$6548,84 doublet are blended in the prism mode. However, owing to the larger SNR and the increased resolution at the red end of the spectrum, we are able to fit a triple Gaussian to this spectral feature, as described in the text and shown in Figure \ref{fig:deblending H-alpha}. The ratio of [N \textsc{ii}]$\lambda$6584/[N \textsc{ii}]$\lambda$6548 is fixed to the theoretical value of 3.049.
\item $^e$ Blended [S \textsc{ii}]$\lambda$6716,31 doublet
\end{tablenotes}
\label{tab:corrected flux catalogue}
\end{table*}

\begin{table*}
\centering
\caption{We present the emission line fluxes without attenuation-correction for the four REBELS galaxies where H$\alpha$ is beyond the wavelength range of NIRSpec, with the same format as in Table \ref{tab:corrected flux catalogue}.}
\begin{tabular}{lcccccccc}
\hline
Galaxy & [O \textsc{ii}] & [Ne \textsc{iii}]3869 & H$\epsilon$ & H$\delta$ & H$\gamma$ & H$\beta$ & [O \textsc{iii}]5007 & [O \textsc{i}]6302 \\ 
\hline
REBELS-12 & 7.1 $\pm$ 0.6 & 2.9 $\pm$ 0.9 & < 1.9 & < 1.4 & < 2.9 & 6.1 $\pm$ 0.6 & 38 $\pm$ 1 & - \\ 
REBELS-14 & 11 $\pm$ 1 & 5.8 $\pm$ 0.7 & 2.6 $\pm$ 0.7 & 2.3 $\pm$ 0.7 & 4.6 $\pm$ 0.7 & 11 $\pm$ 1 & 78 $\pm$ 1 & - \\ 
REBELS-18 & 7.9 $\pm$ 0.4 & 1.9 $\pm$ 0.5 & < 1.2 & < 1.9 & 2.2 $\pm$ 0.5 & 4.3 $\pm$ 0.4 & 21 $\pm$ 0 & - \\ 
REBELS-25 & 6.2 $\pm$ 0.7 & 2.5 $\pm$ 0.6 & < 14 & < 0.19 & 3.1 $\pm$ 0.6 & 5.2 $\pm$ 0.7 & 21 $\pm$ 1 & < 1.3 \\ 
\hline
\end{tabular}
\label{tab:non-corrected flux catalogue}
\end{table*}

\section{Ionised gas properties}
\label{sec:ionised gas properties}
\begin{table*}
\def\arraystretch{1.25}
\centering
\caption{Summary of the properties derived from the NIRSpec/IFU data analysed in this work. Columns show the nebular attenuation ($A_V$), [O \textsc{iii}]$\lambda$5007 EW, O32 ratio, ionisation parameter ($\log U$), oxygen abundance ($12 + \log (\mathrm{O/H})$), stellar mass ($\log (M_*/\mathrm{M_{\odot}})$, Stefanon et al. in prep) and star formation rates from the observed H$\beta$ luminosity (SFR$_{\mathrm{H\beta}}$). For the four sources at $z\gtrsim7$, the O32 ratios given are upper limits since no attenuation-correction is applied.}
\begin{tabular}{lcccccccc}
\hline
Galaxy & $A_{\mathrm{V,neb}}$ (mag) & EW$_{\mathrm{[O \textsc{iii}]}5007}$ (\AA) & O32 & $\log U$ & $12+\log (\mathrm{O/H})$ & $\log M_*/\mathrm{M_{\odot}}$ & SFR$_{H\beta}$ (M$_{\odot}$ yr$^{-1}$) \\ 
\hline
REBELS-05 & 0.95 $\pm$ 0.42 & 372 $\pm$ 44 & 2.43 $\pm$ 0.31 & $-$2.37 $\pm$ 0.04 & 8.51 $\pm$ 0.16 & $9.42^{+0.10}_{-0.10}$ & 137 $\pm$ 16 \\ 
REBELS-08 & 1.21 $\pm$ 0.51 & 372 $\pm$ 95 & 2.35 $\pm$ 0.30 & $-$2.39 $\pm$ 0.04 & 8.22 $\pm$ 0.22 & $9.33^{+0.07}_{-0.06}$ & 143 $\pm$ 21 \\ 
REBELS-12 & - & 387 $\pm$ 40 & $<5.31$ & $<-2.08$ & 8.23 $\pm$ 0.13 & $9.54^{+0.03}_{-0.04}$ & $>26$ \\ 
REBELS-14 & - & 616 $\pm$ 52 & $<7.16$ & $<-1.97$ & 7.90 $\pm$ 0.12 $^\mathrm{a}$ & $9.24^{+0.07}_{-0.06}$ & $>45$ \\ 
REBELS-15 & 0.67 $\pm$ 0.32 & 725 $\pm$ 92 & 9.38 $\pm$ 1.20 & -1.87 $\pm$ 0.07 & 7.78 $\pm$ 0.30 & $9.31^{+0.02}_{-0.01}$ & 204 $\pm$ 18 \\ 
REBELS-18 & - & 195 $\pm$ 27 & $<2.67$ & $<-2.34$ & 8.50 $\pm$ 0.13 & $9.71^{+0.06}_{-0.04}$ & $>21$  \\ 
REBELS-25 & - & 328 $\pm$ 19 & $<3.42$ & $<-2.25$ & 8.62 $\pm$ 0.17 & $9.30^{+0.12}_{-0.14}$ & $>22$ \\ 
REBELS-29 & 0.34 $\pm$ 0.40 & 243 $\pm$ 22 & 3.26 $\pm$ 0.42 & $-$2.27 $\pm$ 0.04 & 8.73 $\pm$ 0.15 & $9.69^{+0.07}_{-0.05}$ & 64 $\pm$ 7 \\ 
REBELS-32 & 0.68 $\pm$ 0.35 & 378 $\pm$ 122 & 3.10 $\pm$ 0.40 & $-$2.28 $\pm$ 0.03 & 8.48 $\pm$ 0.13 & $9.57^{+0.10}_{-0.08}$ & 104 $\pm$ 10 \\ 
REBELS-34 & 1.06 $\pm$ 0.74 & 229 $\pm$ 53 & 2.18 $\pm$ 0.28 & $-$2.41 $\pm$ 0.05 & 8.33 $\pm$ 0.29 & $9.45^{+0.03}_{-0.02}$ & 68 $\pm$ 14 \\ 
REBELS-38 & 1.14 $\pm$ 0.47 & 276 $\pm$ 45 & 1.88 $\pm$ 0.24 & $-$2.47 $\pm$ 0.03 & 8.28 $\pm$ 0.18 & $9.75^{+0.09}_{-0.06}$ & 159 $\pm$ 21 \\ 
REBELS-39 & 0.62 $\pm$ 0.40 & 599 $\pm$ 62 & 5.05 $\pm$ 0.64 & $-$2.10 $\pm$ 0.05 & 8.02 $\pm$ 0.29 & $9.35^{+0.09}_{-0.08}$ & 167 $\pm$ 19 \\ 

\hline
\end{tabular}
\begin{tablenotes}
\item $^a$ For REBELS-14, the upper and lower branch solutions to the R3 calibration are consistent within the uncertainties with the calibration peak. Therefore, we adopt the average of these solutions as the oxygen abundance and the range between them as the uncertainty.
\end{tablenotes}
\label{tab:derived properties}
\end{table*}

With the detected emission lines, we derive the oxygen abundance ($12+\log(\mathrm{O/H})$), ionisation parameter ($\log U$), and star formation rate (SFR) of each galaxy. For these calculations, we adopt the following standard definitions for line ratios:

\begin{equation}
\mathrm{R3}=\mathrm{[O \textsc{iii}]\lambda 5007/ H\beta}
\end{equation}
\begin{equation}
\mathrm{N2}=\mathrm{[N \textsc{ii}]\lambda 6584/ H\alpha}
\end{equation}
\begin{equation}
\mathrm{S2}=\mathrm{[S \textsc{ii}]\lambda 6716,31/ H\alpha}
\end{equation}
\begin{equation}
\mathrm{R23}=\frac{\mathrm{[O \textsc{iii}]\lambda 4959,5007 + [O \textsc{ii}]\lambda3727,29}} {\mathrm{H\beta}}
\end{equation}
\begin{equation}
\mathrm{O32}=\mathrm{[O \textsc{iii}]\lambda 5007/[O \textsc{ii}]\lambda3727,29}
\end{equation}
\begin{equation}
\mathrm{Ne3O2}=\mathrm{[Ne \textsc{iii}]\lambda3869/[O \textsc{ii}]\lambda3727,29}
\end{equation}
\begin{equation}
\mathrm{O2}=\mathrm{[O \textsc{ii}]\lambda3727,29/ H\beta}
\end{equation}

For these line ratios, $\mathrm{[O \textsc{ii}]\lambda3727,29}$, $\mathrm{[O \textsc{iii}]\lambda4959,5007}$,  and $\mathrm{[S \textsc{ii}]\lambda6716,31}$  denotes the sum of the corresponding doublets.

\subsection{Ionisation state}
\label{sec:ionisation state}
%Make sure I'm just selecting SF SDSS galaxies!!!
\begin{figure*}
    \centering
    \includegraphics[width=0.97\textwidth]{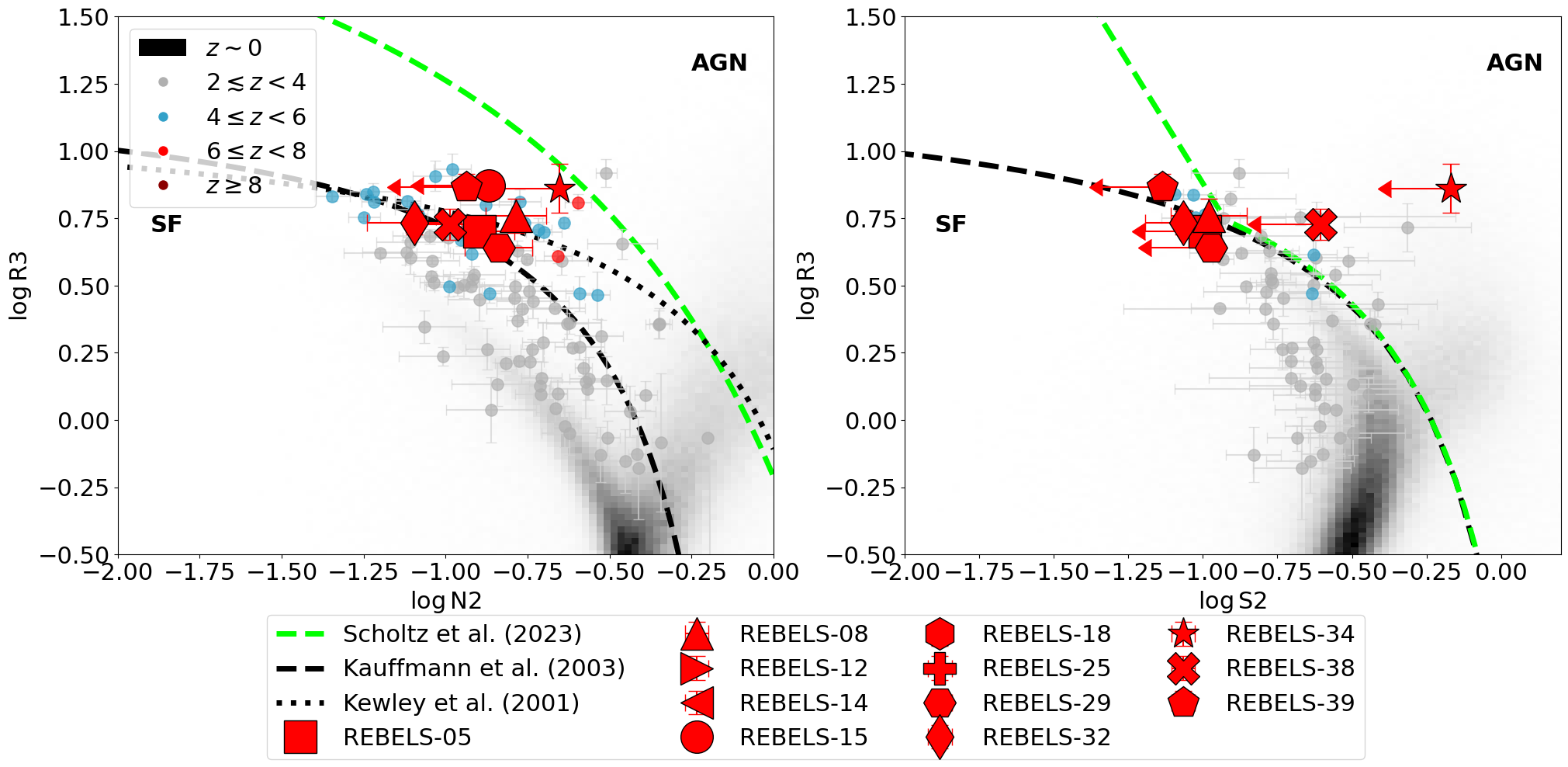}
    \caption{We plot the BPT diagram of R3 against N2 (left panel) and the VO87 diagram of R3 against S2 (right panel) for the the REBELS galaxies where H$\alpha$ is within the NIRSpec wavelength coverage (red markers) and other high-$z$ galaxies from the literature (coloured by redshift, as indicted in the legend in the left panel, taken from JADES data release 3; \protect\citealt{deugenio_jades_2024}, and MOSFIRE; \protect\cite{shapley_mosdef_2015}). In comparison to empirical and theoretical demarcation lines used to distinguish between star-forming (SF) galaxies and those with active galactic nuclei (AGN) (\protect\citealt{kewley_theoretical_2001,kauffmann_host_2003,backhaus_clear_2022, scholtz_jades_2023}), we find that the REBELS sources are mostly consistent with ionisation predominantly by star formation.}
    
    \label{fig:BPT plots}
\end{figure*}

\begin{figure*}
    \centering
    \includegraphics[width=0.97\textwidth]{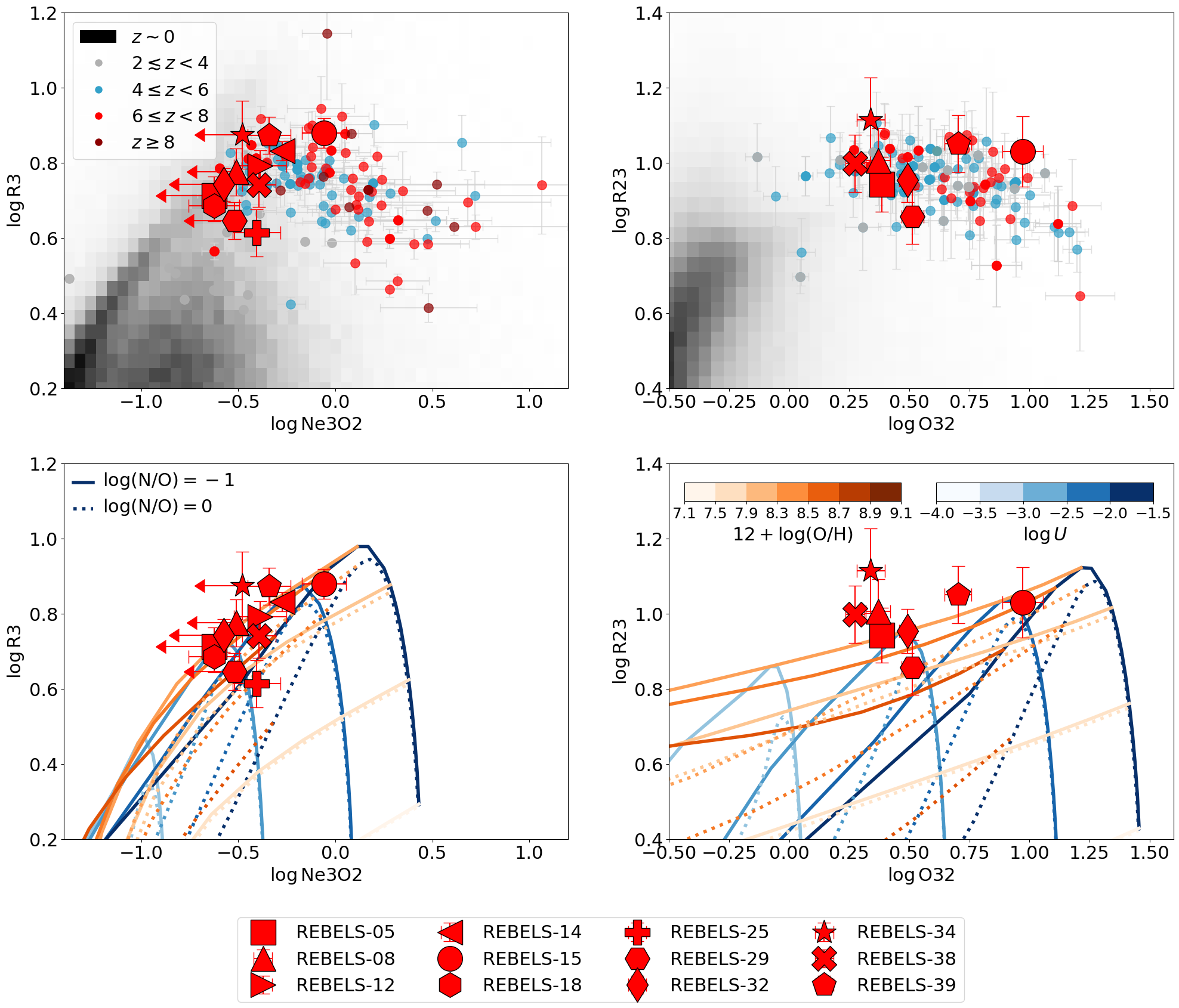}
    \caption{We present two emission line ratio diagrams that can be used to trace the ionisation and metallicity of galaxies. In the left panels, we show R3 versus Ne3O2 using emission line ratios without attenuation correction for all 12 of the REBELS galaxies analysed in this work, as these lines are close in wavelength. For five of the REBELS galaxies, the [Ne \textsc{iii}] line is detected at $<3\sigma$, and we therefore show these Ne3O2 ratios as $3\sigma$ upper limits. In the right panels, we plot R23 versus O32 for only the eight REBELS galaxies where attenuation-corrected fluxes have been derived (see text). In the top panels, we compare the REBELS galaxies to literature samples across various redshifts, including SDSS galaxies at $z\sim0$, MOSDEF (\protect\citealt{shapley_mosdef_2015}) galaxies at $z\sim2-4$, and sources from the PRIMAL survey (\protect\citealt{heintz_jwst-primal_2024}, full references given therein) for $z>4$. In the bottom panels, we compare the position of the REBELS galaxies on these diagrams with \texttt{CLOUDY} photoionisation model grids at different values of $12+\log(\mathrm{O/H})$, $\log U$ and $\log (\mathrm{N/O})$. We find that most of the REBELS galaxies have O32 and Ne3O2 ratios towards the lower end of those $z>6$ galaxies in existing surveys, indicating lower ionisation parameters and higher metallicities.
    }
    \label{fig:OHNO plots}
\end{figure*}

Before deriving the ISM properties of these galaxies, we first investigate their source of ionisation, in particular to examine potential contamination from active galactic nuclei (AGN). Emission line diagnostic diagrams, such as BPT (\citealt{baldwin_classification_1981}), VO87 (\citealt{veilleux_spectral_1987}), and `OHNO' or ionisation-metallicity (\citealt{backhaus_clear_2022}) diagrams are often used to investigate the ionisation state of galaxies and regions of ionised gas. For example, in the local Universe, gas predominantly ionised by star formation typically follows well-defined sequences in these diagrams (e.g., \citealt{kewley_theoretical_2001,kauffmann_host_2003}), whereas gas ionised by harder ionising sources, like AGNs, typically shows enhanced N2 and S2 ratios. We plot the BPT and VO87 diagnostic diagrams in Figure \ref{fig:BPT plots} for the 12 REBELS galaxies, as well as galaxies in the literature at $z\sim0$ from the SDSS DR8 catalog, $z\sim2-3$ from MOSDEF (\citealt{shapley_mosdef_2015-1}), and sources from JADES data release 3 catalogues (\citealt{deugenio_jades_2024}).

For nearby galaxies, the empirical demarcation lines from \cite{kewley_theoretical_2001} and \cite{kauffmann_host_2003} are typically used to distinguish between AGN, star-forming, and composite galaxies in the BPT and VO87 diagrams. However, at high redshifts, these demarcations may no longer be applicable since studies have shown that galaxies at high redshift do not lie on the same loci of nebular line ratios as local galaxies (e.g., \citealt{shapley_chemical_2005,liu_metallicities_2008,steidel_strong_2014,strom_nebular_2017}). This offset from the local sequence at high-$z$ is likely driven by a difference in ionisation properties, such as lower metallicities and harder ionising fields. For this reason, a recent study (\citealt{scholtz_jades_2023}) has defined new, conservative demarcation lines for the BPT and VO87 diagrams based on star-forming models from \cite{gutkin_modelling_2016}. However, [N \textsc{ii}] and [S \textsc{ii}] are largely undetected at $z>4$, even in large-scale surveys like JADES and CEERS, meaning statistics are limited in testing these demarcations. In a sample of 38 galaxies at $5\leq z <6.5$ from the CEERS survey studied in \cite{sanders_excitation_2023}, there is only one detection of [S \textsc{ii}], and six detections of [N \textsc{ii}]. From the JADES survey, as analysed in \cite{cameron_jades_2023}, there are only three detections (in a sample of 20 galaxies at $5.5<z<6.9$) of [S \textsc{ii}] and no detections of [N \textsc{ii}] (in a sample of 18 galaxies at $5.5<z<6.9$), with [N \textsc{ii}] also undetected in a stack of all 18 spectra.  In contrast, with just 30 mins of exposure per source of the REBELS galaxies, [N \textsc{ii}]$\lambda6584$ is detected at $>3\sigma$ for five out of eight galaxies with the correct wavelength coverage in the REBELS/\textit{JWST} sample (tentatively detected at $\gtrsim2\sigma$ for the remaining three), and [S \textsc{ii}] is detected at $>3\sigma$ in two out of seven (tentatively detected at $\gtrsim2\sigma$ for an additional three sources). However, we again emphasise the caveat that the [N \textsc{ii}] line is blended with H$\alpha$, increasing the uncertainty of the N2 ratio derived from the prism spectra.

From these BPT and VO87 diagnostic plots, we see no strong evidence of AGN activity within the REBELS sources, since all detections lie below the conservative \cite{scholtz_jades_2023} demarcation line. For REBELS-34 and REBELS-38, whilst their 3$\sigma$ upper limits on S2 lie above this demarcation in the VO87 diagram, they lie below it on the BPT diagram. There are, however, some potential AGN candidates (REBELS-15, REBELS-34, and REBELS-39) that show slightly elevated R3 ratios in comparison to the \cite{kewley_theoretical_2001} and \cite{kauffmann_host_2003} demarcations within the sample. However, these sources, and the use of these diagnostics to identify AGN at high-$z$, would require further study to confirm or deny the presence of AGN. Follow-up \textit{JWST} Cycle 3 observations (GO-6480, PI S. Schouws) of galaxies from the REBELS sample may shed further light on the nature of these sources by obtaining high spatial resolution imaging with grism spectroscopy, which will enable, for example, a search for broad emission line features which are also characteristic of AGN activity.

In Figure \ref{fig:OHNO plots}, we plot two different ionisation-metallicity diagnostic diagrams ---  R3 versus Ne3O2, and R23 versus O32 --- for the REBELS galaxies and the aforementioned literature samples spanning $z=0$ to $z\gtrsim8$, with the addition of $z>4$ sources from the PRIMAL survey (\citealt{heintz_jwst-primal_2024}), which includes sources from CEERS (ERS-1345; \citealt{finkelstein_ceers_2023}), GLASS-DDT (DDT-2756 \citealt{treu_glass-jwst_2022}), JADES (GTO-1180, 1210, GO-3215; \citealt{bunker_jades_2023}; \citealt{eisenstein_overview_2023}), and UNCOVER (GO-2561; \citealt{bezanson_jwst_2024}). Whilst we find similar R3 and R23 ratios to other $z\gtrsim6$ galaxies, the REBELS sources show lower O32 and Ne3O2 ratios (and also O2 ratios). Recent \textit{JWST} observations find O32 ratios typically between 7 to 31, Ne3O2 $\sim 1$-2 and O2 $\gtrsim5$ from reionisation-era galaxies (\citealt{rhoads_finding_2023,tang_jwstnirspec_2023,sanders_excitation_2023,mascia_closing_2023,roberts-borsani_between_2024}). In low-redshift galaxies, such high O32, Ne3O2, and O2 ratios are mostly associated with extreme [O \textsc{iii}] emitters (\citealt{tang_mmtmmirs_2019}) and/or those with high ionisation parameter and low metallicity (\citealt{strom_measuring_2018,papovich_clear_2022}). In contrast, all but two of the REBELS galaxies have O32 ratios $<5$, with Ne3O2 and O2 ratios $\lesssim0.8$ across the entire sample. This suggests that the REBELS galaxies probe different ISM conditions compared to other existing spectroscopic surveys at $z > 6$, likely representing more evolved, metal-rich galaxies with less extreme ionising conditions. In addition, existing large \textit{JWST} surveys may be more biased toward galaxies with high O32 values and high [O \textsc{iii}] EWs. In contrast, the [O \textsc{iii}] EWs of our sample (listed in Table \ref{tab:derived properties}) are more consistent with the median [O \textsc{iii}] EW at $z \sim 7$ of $\sim 450$ \AA\ (\citealt{labbe_spectral_2013}), suggesting that the REBELS galaxies may provide a more representative view of the general galaxy population at these redshifts, rather than being skewed toward extreme line emitters.

Assuming these sources are predominantly ionised by star formation, we compare with photoionisation models from \texttt{HII-CHI-Mistry} (\citealt{perez-montero_deriving_2014}). These models used the synthesis spectral code CLOUDY v13.03 (\citealt{ferland_2013_2013}) using POPSTAR (\citealt{molla_popstar_2009}) stellar evolutionary models assuming an instantaneous burst with an age of 1 Myr with an initial mass function from \citealt{chabrier_galactic_2003}. They varied the ionisation parameter between $-4.00\leq \log(U) \leq -1.50$ in steps of 0.25 dex, the oxygen abundance between $7.1 \leq 12+\log(\mathrm{O/H}) \leq 9.1$ in steps of 0.1 dex, and considered variations in the N/O ratio between $0.0 \leq \log(\mathrm{N/O}) \leq -2.0$ in steps of 0.125 dex, thus totalling 3927 models. We plot grids of these models for N/O values of $-$1.0 and 0.0 in the bottom panels of Figure \ref{fig:BPT plots} and Figure \ref{fig:OHNO plots}. Here, we see that the REBELS galaxies lie in a parameter space with ionisation parameters $\log U \lesssim-2$, but with multiple solutions for the oxygen abundance.

To assess the ISM properties of these sources, we first further investigate their ionisation parameters. The ionisation parameter, $\log U$, is commonly defined as the ratio of ionising photon density to hydrogen density and can be used to indicate the ionisation state (e.g., \citealt{tarter_interaction_1969}). It is often derived from the O32 ratio, as this ratio represents the relative abundance between doubly-ionised oxygen to singly ionised oxygen (\citealt{strom_measuring_2018,kewley_theoretical_2019,papovich_clear_2022}). To estimate $\log U$ from O32, we use the calibrations empirically derived in \cite{papovich_clear_2022} from a sample of galaxies at $z\sim 1.1-2.3$. We list these derived values in Table \ref{tab:derived properties}. We find a range of $\log U$ from $-$2.5 to $-$1.9, and we find that these values are consistent within the uncertainties when derived using the calibrations from \cite{diaz_chemical_2000} and \cite{kewley_theoretical_2019}, and when using the aforementioned \texttt{HII-CHI-Mistry} code. However, we note that these calibrations are based on lower redshift sources, and may therefore need some adjustment at $z>6$.

Having found that these galaxies exhibit lower O32 ratios, and therefore lower ionisation parameters, than many other high-$z$ sources in the literature, we may therefore expect to find higher metallicities in comparison to the literature samples (e.g. \citealt{sanders_mosdef_2021}). We describe the methods used to derive the metallicities of these galaxies below.

\subsection{Gas-phase metallicity}
\label{sec:metallicity measurements}

To derive the gas-phase metallicity, as traced through the oxygen abundance,  the direct method is generally preferred (e.g. \citealt{peimbert_nebular_2017,kewley_theoretical_2019}). The direct method is dependent on the detection of temperature-sensitive auroral lines, the most common of which is [O \textsc{iii}]$\lambda$4363. However, the resolution of the NIRSpec prism is insufficient to resolve this line from the neighbouring H$\gamma$ line (and the [FeII]$\lambda$4360 line, which may be more dominant for these galaxies, see e.g. \citealt{curti_new_2017,shapley_aurora_2024}) and other auroral lines within the NIRSpec wavelength range are undetected for our sample. This is perhaps not surprising, since auroral lines are difficult to detect in massive galaxies even in the low redshift Universe, as their brightness is inversely proportional to the gas-phase metallicity. We therefore use strong line calibrations to derive oxygen abundances for the 12 galaxies in our sample.

A variety of both empirical and theoretical strong line metallicity calibrations exist within the literature. Prior to the launch of JWST, empirical calibrations were based on high-$z$ analogs in the local Universe (e.g. \citealt{maiolino_amaze_2008,bian_direct_2021,curti_mass-metallicity_2020,nakajima_empress_2022}). However, the use of calibrations based on local analogs may be unreliable for high-$z$ galaxies due to evolving ISM conditions, which can alter the relationship between strong-line ratios and gas-phase metallicity, potentially skewing the derived oxygen abundances. More recently, thanks to the growing number of auroral line detections at $z>2$, new high-$z$ calibrations have now been derived and tested (\citealt{sanders_direct_2024}). The strong-line calibrations derived in \cite{sanders_direct_2024}, based on 46 auroral line detections at $1.4 \leq z \leq 8.7$, are calibrated for $7.0 \leq 12+\log(\mathrm{O/H}) \leq 8.4$.  However recent auroral line detections at $z \sim 2-3$ up to $12+\log(\mathrm{O/H}) \sim 8.5$ from the MARTA survey find good agreement with these calibrations (Cataldi et el. in prep.)\, as do absorption-line based gas-phase metallicities at $z\sim2-4$ up to $12+\log(\mathrm{O/H})\sim8.7$ (\citealt{schady_comparing_2024})\footnote{In Table D1 of \cite{schady_comparing_2024}, the low-branch solutions using the R3 and R23 \cite{sanders_direct_2024} calibrations are listed. However, based on the emission line fluxes reported therein, we re-derive the metallicities, using the O32 ratio to choose between the low-branch and extrapolated high-branch solutions, and find better agreement with the absorption line-based metallicities, increasing our confidence in extrapolating the \cite{sanders_direct_2024} calibrations to higher oxygen abundances.}. We therefore adopt these calibrations to derive the fiducial metallicities for our sample. We also provide a comparison with other calibrations (\citealt{bian_direct_2021,nakajima_empress_2022,laseter_jades_2024}) and with the \texttt{genesis-metallicity} calibrator (\citealt{langeroodi_genesis-metallicity_2024}) in Appendix \ref{sec:appendix_metallicity}, and we detail the caveats in Section \ref{sec:caveats}, below. 

Of the different strong-line ratios found to correlate with metallicity, the R23 index is generally found to show the least scatter, and is used often in the literature in both the nearby Universe (e.g. see \citealt{tremonti_origin_2004} and references therein) and now out to high redshifts (e.g. \citealt{nakajima_jwst_2023}). We therefore use this ratio where possible. However, for the four galaxies where H$\alpha$ is not detected and a reddening-correction is not applied, we instead use the R3 ratio, which is less impacted by interstellar reddening and flux calibration uncertainties. A limitation of using the R23 and R3 ratios as metallicity callibrators is that they typically yield two different metallicity estimates for the same ratio. Following, for example, \cite{kewley_metallicity_2008,nakajima_jwst_2023,curti_jades_2024}; and \cite{sarkar_unveiling_2024}, we use the O32 and Ne3O2 ratio or upper limit to select the branch of the R23 and R3 relations. As indicated from the lower O32 ratios discussed in Section \ref{sec:ionisation state},  the high branch solution is preferred for all but one of the REBELS galaxies. For the remaining galaxy, REBELS-14, both solutions are consistent with the peak of the R3 relation at $12+\log(\mathrm{O/H})\sim 7.92$, and we therefore take the two solutions as upper and lower bounds for its oxygen abundance, as in \citealt{nakajima_jwst_2023}.

We note that for three galaxies (REBELS-15, REBELS-34 and REBELS-39), there is no solution for the R3 or R23 calibration due to their high R3 and R23 ratios. For these galaxies, we use the O32 ratio to estimate the oxygen abundance. 

The oxygen abundances for the sample are listed in Table \ref{tab:derived properties}. Notably, all galaxies are found to have $Z_{g}>0.1 Z_{\odot}$, and we find some near-solar and potentially even super-solar oxygen abundances within the sample. However, it is important to treat these absolute values with caution due to the significant systematic uncertainties associated with strong-line calibrations. For consistency, we focus on comparing our results with other high-$z$ galaxies in the literature where we have applied the same calibration methodology. We discuss the various caveats in our derived oxygen abundances in more detail below.

\subsubsection{Caveats}
\label{sec:caveats}

Even in the local Universe, where metallicity calibrations are widely studied,  metallicities estimated through different strong line ratios and calibrations can vary by as much as $\sim0.6$ dex (\citealt{kewley_metallicity_2008,moustakas_optical_2010,rowland_pre-supernova_2024}). At high-$z$, where the number of auroral line detections is lower and ionising conditions are less well-understood, discrepancies are likely to be even greater. The \cite{sanders_direct_2024} calibrations adopted in this paper are based on a sample size of only 46 galaxies, over half of which are at $z<5$. It is clear that a more significant sample of auroral line detections at high-$z$ is essential for improving these calibrations, although we note that auroral line detections reported since the publication of \cite{sanders_direct_2024} have found good agreement with these calibrations (e.g. \citealt{laseter_jades_2024}), even out to $z=10.17$ (\citealt{hsiao_first_2024}).

The main source of uncertainty with our use of these calibrations is that they are only calibrated up to $12+\log(\mathrm{O/H})\sim8.4$, meaning that for five galaxies in our sample, we are extrapolating. The same issue is present in, for example, the \cite{bian_direct_2021} and \cite{nakajima_empress_2022} calibrations. The general expectation has been that high-$z$ galaxies would be metal-poor, meaning that most calibrations based on local analogs have focused on extremely meal-poor galaxies. Recent findings of a mature, metal-rich galaxy at $z=6.7$ (\citealt{shapley_aurora_2024}), as well as the REBELS galaxies studied here, highlight that it is now necessary to extend these calibrations to higher metallicities. 

Additionally, there are also caveats to consider in the choice of line ratio used to derive the oxygen abundance. Whilst the R23 and R3 ratios are found to be the most accurate metallicity indicators at high-$z$ (\citealt{nakajima_jwst_2023,laseter_jades_2024}), they span only a narrow dynamic range at these redshifts ($\sim 0.4$ dex in Figures \ref{fig:BPT plots} and \ref{fig:OHNO plots}; see also \citealt{sanders_mosdef_2021}), mostly scattered around the turnover regime between the ‘upper’ and ‘lower’ branches of these indices. It could therefore be argued that more linear calibrations, such as O2, O32, and Ne3O2, may provide more reliable extrapolations. However, they have a greater intrinsic scatter in \cite{sanders_direct_2024} and recent studies have found that these ratios may not be good indicators of metallicity at these redshifts (e.g. \citealt{laseter_jades_2024}). They are also generally not recommended at lower redshifts due to their strong, primary dependence on the ionisation parameter (e.g. \citealt{kewley_using_2002,patricio_testing_2018}). Furthermore, the ionisation-metallicity relation is not constant across all redshifts, such that O2, O32, and Ne3O2 indices calibrated at a lower redshift tend to underestimate the metallicity for higher redshift sources (\citealt{sanders_mosdef_2021,garg_theoretical_2024,sanders_direct_2024,laseter_jades_2024}). Indeed, \cite{tang_mmtmmirs_2019} and \cite{reddy_jwstnirspec_2023} find that the specific star formation rate (sSFR) and gas density may play more central roles than metallicity in modulating $\log U$ (and hence indices like O2, O32 and Ne3O2) at these redshifts. These indices tend to predict lower oxygen abundances for the 12 REBELS galaxies, with a maximum $12+\log(\mathrm{O/H})$ of 8.4 (see Appendix \ref{sec:appendix_metallicity}). We note that the N2 ratio also shows a monotonic relation with metallicity, and whilst it is has not yet been calibrated with high-$z$ auroral line detections, the calibrations based on local analogs (\citealt{nakajima_empress_2022}) are more consistent with the fiducial metallicities of the REBELS sample adopted in this work (see Table \ref{tab:full sample nakajima}).

We have elected to use the R23 and R3 ratios for our fiducial metallicity estimates where possible (but use O32 for the three sources where the R3 and R23 calibrations have no real solution). We list the oxygen abundances derived using O32, Ne3O2, $\hat{R}$ (defined in \citealt{laseter_jades_2024}), N2, and other indices for all the galaxies in our sample in Appendix \ref{sec:appendix_metallicity}, and we also re-analyse the MZR and FMR based on these values therein. Due to the scatter in our sample, and in the literature, using these ratios does not significantly change the derived MZR and FMR, and we find that the REBELS galaxies still have oxygen abundances as much as $\sim 0.4$ dex higher than other galaxies at $6<z<8$. Therefore, whilst it is important to treat the absolute values with caution, it is clear that the REBELS galaxies are more metal-rich compared to most EoR galaxies in other existing spectroscopic surveys.

\subsection{Star formation rate}
\label{sec:star formation rates}
Following, for example, \cite{heintz_dilution_2023} and \cite{nakajima_jwst_2023}, we use H$\beta$ as a SFR tracer in lieu of the commonly used H$\alpha$ luminosity, as an indicator for ongoing ($\sim$10 Myr) star formation activity. This choice is necessitated by the spectral coverage limitation of NIRSpec, which does not extend to H$\alpha$ at redshifts $z\gtrsim 7$. This use of H$\beta$ therefore allows us to maintain consistency in SFR measurements across our sample, and with measurements at high-$z$ in the literature. The SFR is derived from the H$\beta$ luminosity, $L_{\mathrm{H\beta}}$, assuming a \cite{kroupa_variation_2001} IMF following \cite{kennicutt_global_1998}:

\begin{equation}
\mathrm{SFR_{H\beta}} (\mathrm{M_\odot yr^{-1}})= 5.37 \times 10^{-42} \times L_{\mathrm{H\beta}} (\mathrm{erg s^{-1}}) \times 2.86.
\end{equation}

For the galaxies where we do not have attenuation-corrected H$\beta$ luminosities, we determine lower limits to the SFR from the observed H$\beta$ emission. We add that a comparison of the SFR of this sample from different tracers will be the focus of a subsequent paper (Fisher et al. in prep ). When comparing with other galaxies in the literature, we apply a correction factor to their reported SFRs if other IMFs are used, so that the resulting SFRs are consistent with using a \cite{kroupa_variation_2001} IMF.

\section{SED fitting}
\label{sec:sed fitting}

We derive stellar masses for the REBELS galaxies using the \texttt{BAGPIPES} spectral energy distribution (SED) fitting code (2016 version). We briefly outline our method here, with full details given in Stefanon et al. (in prep). We use the \cite{bruzual_stellar_2003} stellar population models, which assume a \cite{kroupa_variation_2001} IMF. Nebular emission is modelled using CLOUDY (\citealt{ferland_2017_2017}), assuming a spherical constant-density gas distribution with $n(H)=100$ cm$^{-3}$, matching the metallicity of the stellar component. The metallicity priors for the SED fitting were Gaussian, centred on the metallicity estimates derived in this work, with the standard deviation reflecting the associated uncertainties. The redshifts were fixed to the value measured from the [C \textsc{ii}]158$\mu$m emission, accounting for small systematic shifts in the NIRSpec wavelengths (see Stefanon et al. in prep). We adopted a \cite{calzetti_dust_2000} attenuation curve, allowing the stellar dust attenuation ($A_{\mathrm{V}}$, i.e. different to the nebular attenuation determined using the Balmer decrement in Section \ref{sec:reddening correction}) to vary freely in the range 0$-$3 mag under a flat prior. For the star-formation history (SFH), we employed a non-parametric model with a continuity prior, designed to capture both recent bursts and early assembly phases. Other SFH parameterizations, including a constant+burst model, were also explored in Stefanon et al. (in prep.).

With this SED fitting, we find a range in stellar masses of $\log (M_*/\mathrm{M_\odot})\sim 9.2-9.8$ (presented in Table \ref{tab:derived properties}). These stellar masses are slightly lower than those found for the same galaxies from non-parametric SED fitting in \cite{topping_alma_2022}. A full comparison of these different results will be presented in Stefanon et al. (in prep). The continuum attenuation values derived from the SED fitting are lower than the nebular attenuation derived from the Balmer decrement, as expected for star-forming galaxies (e.g. \citealt{calzetti_dust_1994}).  

While the non-parametric approach provides flexibility in reconstructing the assembly history, the stellar mass measurements of these sources can still be impacted by the `outshining' effect, where young, luminous stellar populations dominate the observed light, potentially causing a systematic underestimation of the stellar masses (e.g. \citealt{gimenez-arteaga_spatially_2023}). However, the impact of outshining is expected to diminish at higher stellar masses, as discussed in \cite{lines_jwst_2024}. Dust obscuration can also present challenges when deriving stellar masses and other stellar population parameters. For the REBELS sources analysed here, it is likely that the observed light in the rest-frame UV/optical (as traced by these \emph{JWST} observations) is primarily emitted by less-obscured regions. However, there may be fully obscured star-forming regions that do not emit at all in these wavelengths and are only detectable in the infrared. This results in an incomplete picture of the total stellar mass if only unobscured regions are traced. Such obscured regions are evident in some REBELS sources, where spatial offsets between UV/optical and infrared emission suggest the presence of highly obscured star formation (e.g., \citealt{hygate_alma_2023, rowland_rebels-25_2024}).  Future spatially resolved SED analyses, including comparisons with high-resolution ALMA FIR data, will be critical for mitigating such uncertainties and providing more accurate estimates of stellar mass distributions within these galaxies.

\section{Discussions}
\label{sec:discussions}

\begin{figure*}
    \centering
    \includegraphics[width=0.9\textwidth]{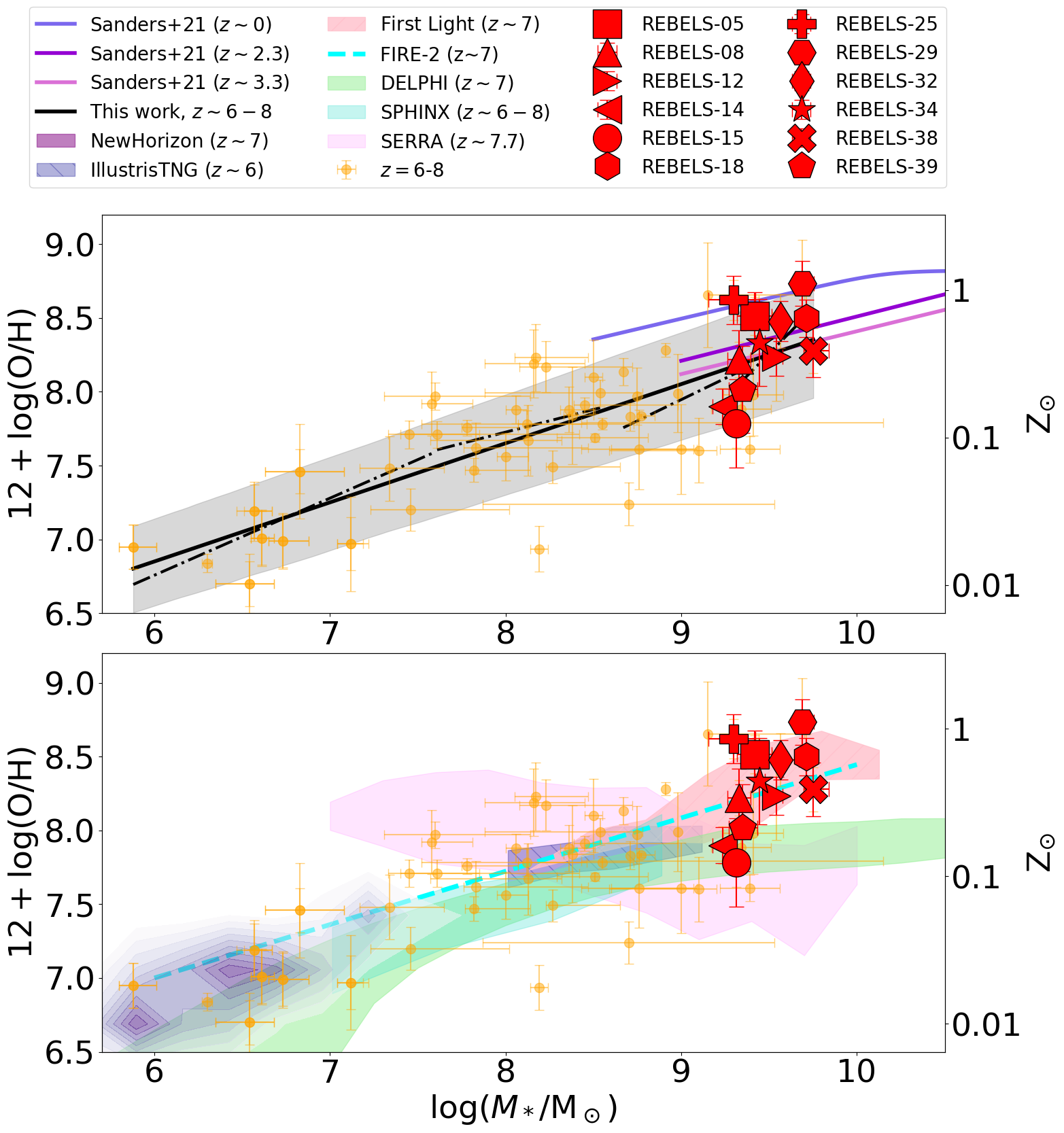}
    
    \caption{We show the MZR at $z=6$-$8$ for the REBELS galaxies and sources from the literature (\protect\citealt{nakajima_jwst_2023,chemerynska_extreme_2024}). In the top panel, the best-fit MZR is indicated by the solid black line, with the $1\sigma$ uncertainty shown with the grey shaded area. We also plot the MZR at $z\sim0$, $z\sim2.3$, and $z\sim3.3$ from \protect\cite{sanders_mosdef_2021} to investigate the redshift evolution of this scaling relation. The black dash-dot lines show the best-fit MZR in four mass bins, each with 15 galaxies, across this 4 dex range in stellar mass. In the bottom panel, we compare the MZR to theoretical predictions from \texttt{NEWHORIZON} (\protect\citealt{dubois_introducing_2021}), TNG100 (\protect\citealt{torrey_evolution_2019}), First Light (\protect\citealt{nakazato_simulations_2023}), FIRE-2 (\protect\citealt{marszewski_high-redshift_2024}), DELPHI (Mauerhofer et al. in prep), and SPHINX (\protect\citealt{katz_sphinx_2023}). For FIRE-2, we plot the best-fit MZR from the `Slope and Normalization Evolution' model at $z=7$. For FirstLight, TNG100, DELPHI and SPHINX we show the 16th-84th percentile distribution of these predictions by the coloured regions.} 
    \label{fig:MZR}
\end{figure*}

\subsection{The Mass-Metallicity Relation (MZR)}
\label{sec:mzr}

With these observations of the REBELS galaxies, we are able to increase the sampling at the high-mass end of the MZR in star-forming galaxies during the EoR, since typical large-scale \textit{JWST} surveys mainly contain galaxies with $\log (M_*/\mathrm{M_{\odot}}) < 9$. We make comparisons with observations at lower stellar mass, including a sample of galaxies from CEERS (\citealt{finkelstein_ceers_2023}), GLASS (\citealt{roberts-borsani_between_2024}), some ERO programs collected and analysed in \cite{nakajima_jwst_2023}, and from the UNCOVER (\citealt{bezanson_jwst_2024}) program (analysed in \citealt{chemerynska_extreme_2024}). Selecting galaxies only in the redshift range of $6<z<8$, we have a sample of 49 galaxies with publicly available stellar masses, SFRs, and the relevant emission line ratios to derive $12+\log(\mathrm{O/H})$ with a consistent methodology using the \cite{sanders_direct_2024} calibrations (i.e. [O \textsc{iii}], H$\beta$, [O \textsc{ii}] and/or an additional Balmer line). For the sources in \cite{chemerynska_extreme_2024}, the R3 \cite{sanders_direct_2024} calibrations were already used to determine the oxygen abundance. For the \cite{nakajima_jwst_2023} sources, the \cite{nakajima_empress_2022} calibrations were used. To be consistent, we recalibrate the \cite{nakajima_jwst_2023} sources using the \cite{sanders_direct_2024} calibrations, following the methodology outlined in Section \ref{sec:metallicity measurements}. However, we note that the main findings of this analysis are consistent even if we use the \cite{nakajima_empress_2022} R3 or R23 calibrations for all the sources.

We plot the MZR for the 12 REBELS galaxies, and the literature sources at $z\sim 6-8$, in Figure \ref{fig:MZR}. Here, we see that \textit{JWST} observations in this redshift range span around 4 dex of stellar mass, and the addition of the REBELS galaxies more than doubles the number of galaxies at $z>6$ with $M_*>10^{9} \mathrm{M_{\odot}}$ in comparison to other recent MZR studies (\citealt{nakajima_jwst_2023,curti_jades_2023,heintz_dilution_2023,sarkar_unveiling_2024}).

Following \cite{sanders_mosdef_2021}, we fit the MZR using:

\begin{equation}
12+\log (\mathrm{O/H}) = \gamma \log (\frac{M_*}{10^{10}\mathrm{M_{\odot}}}) + Z_{10}
\label{eq:MZR}
\end{equation}

\noindent and find a best-fit slope with $\gamma = 0.37\pm0.03$ and $Z_{10}=8.30\pm0.04$, with a 1$\sigma$ scatter of 0.26 dex. We plot the best-fit linear relation for the MZR as a solid black line in the top panel of Figure \ref{fig:MZR}. The gray shaded region represents the uncertainty in the best-fit, calculated as the range between the 16th and 84th percentiles from 10,000 Monte Carlo simulations of the posterior distributions of $\gamma$ and $Z_{10}$. In this top panel, we also plot the MZR relation at lower redshifts from \cite{sanders_mosdef_2021} (at $z\sim0$, $z\sim2.3$, and $z\sim3.3$), although we note that these metallicities are calculated with a $\chi^2$-minimisation using O32, Ne3O2, and R3 ratios. We find that the fitted MZR at $z=6$-$8$ is consistent with the MZR reported in \cite{heintz_dilution_2023} based on a different sample selection ($\gamma=0.33$ at $z=7$-$10$) and from \cite{chemerynska_extreme_2024} ($\gamma= 0.39$ at $z=6$-$8$). The latter study was based on a similar sample to the one analysed here (excluding the REBELS sources) but did not apply a consistent metallicity calibration methodology across the full sample, as done in this work. However, these MZR slopes are steeper than those found in \cite{nakajima_jwst_2023,curti_jades_2023}; and \cite{sarkar_unveiling_2024} at $z>4$.  We also find that this best-fit MZR at $z=6$-$8$ is steeper than the lower redshift MZRs reported in \cite{sanders_mosdef_2021}. As aforementioned, there is some caution to be taken in comparing these different fitted MZRs, since different works use different SED fitting parameters and metallicity calibrations. We show in Appendix \ref{sec:appendix_metallicity} that using the O32 ratio for this sample results in slightly lower $\gamma=0.27\pm0.07$ and $Z_{10}=8.17\pm0.08$, but this best-fit is still consistent within the uncertainties.

Whilst most MZR studies at $z>2$ find that oxygen abundances at a given $M_*$ are significantly lower at high-$z$, the REBELS galaxies lie in a similar parameter space as the lower-$z$ (and even $z\sim0$) relations, although we note that there is significant scatter within the sample. This alignment could be attributed to the observed flattening of the MZR at high masses at $z \sim 0$, with a characteristic turnover mass around $\log(M_0/\mathrm{M_{\odot}}) \sim 10$ (\citealt{curti_mass-metallicity_2020}). This flattening is often attributed either to the advanced stage of chemical evolution in massive galaxies, a phenomenon known as “chemical downsizing” (\citealt{somerville_physical_2015}), or to the enhanced ability of massive galaxies to retain metals due to their deep gravitational wells. In contrast, lower-mass galaxies more readily lose enriched gas via outflows and winds (e.g., \citealt{chisholm_scaling_2015}). The MZR turnover occurs when gas-phase oxygen levels become high enough that a significant fraction of oxygen is locked in low-mass stars. This turnover mass, $M_0$, is roughly where the stellar-to-gas mass ratio is unity, or equivalently, where the gas fraction $f_{\mathrm{gas}} \sim 0.5$ (\citealt{zahid_universal_2014}).  The turnover mass has been found to increase with increasing redshift out to $z \sim 1.5$ (\citealt{zahid_universal_2014}), with indications that this is also true out to $z\sim2.3$ (\citealt{sanders_mosdef_2021}). To tentatively assess whether there is any flattening of the MZR in this sample of $z=6$-$8$ galaxies, we create four bins of 15 galaxies according to their stellar mass, and recalculate the MZR in each bin according to Equation \ref{eq:MZR}. The slopes for each mass bin are indicated by the black dash-dot lines in the top panel of Figure \ref{fig:MZR}. With this sample of $6\lesssim \log({M_*/\mathrm{M_\odot}})\lesssim 10$ galaxies at $z=6$-$8$, we see no evidence of the MZR flattening at the high-mass end, although we note there is significant scatter. Indeed, from the [C \textsc{ii}]-based gas masses and derived gas fractions (\citealt{aravena_alma_2024}, Algera et al. submitted) of the REBELS sample, all the gas fractions are $>0.61$, indicating that, if the flattening of the MZR does persist out to these redshifts, we may not have yet reached the turnover mass ($M_0(z\sim7)>10^{10}\mathrm{M_\odot}$). However, as discussed in Section \ref{sec:sed fitting}, integrated SED fitting can underestimate the derived stellar masses by as much as $\sim 0.5$ dex (e.g. \citealt{gimenez-arteaga_spatially_2023}) due to the `outshining' effect, which may particularly be a problem for the REBELS galaxies, based on the finding that they may be more evolved systems with potentially older stellar populations that are outshone by the emission from brighter, younger stellar populations. Spatially-resolved SED fitting of the REBELS galaxies in subsequent works will be able to assess this effect on the derived slope of the MZR in more detail.

In \cite{chemerynska_extreme_2024}, the scatter in the MZR at $z\sim6$-$8$ is found to increase at lower stellar masses, interpreted as due to more stochastic star formation and ISM enrichment at these low-masses. With the recalibrated sample presented here, we find a relatively consistent scatter of $\sim0.25$ dex across the four mass bins from $\log(M_*/\mathrm{M_{\odot}})=6$ to $10$. This scatter is significantly higher than the $\sim0.1$ dex scatter typically found at low redshifts (e.g. \citealt{andrews_mass-metallicity_2013}), and is consistent with the MZR scatter from the SERRA simulations at $z\sim 7$ (\citealt{pallottini_mass-metallicity_2024}). This scatter is also consistent with the analytic models described in \cite{pallottini_mass-metallicity_2024} with a weak supernova feedback efficiency ($\epsilon_{\rm SN}=1/4$) and a low-amplitude SFR flickering ($\sigma_{\rm SFR}\approx 0.2$), resulting in an intrinsic scatter of $\sim0.25$ dex at $z=3-10$. However, we note that the scatter in the REBELS sample, as well as other literature samples at high-$z$, may be affected by other intrinsic or systematic effects, such as systematics in the metallicity calibrations used, systematics in the stellar mass estimates, contamination from AGN, and effects from galaxy interactions/mergers. %To more robustly assess the scatter in the MZR at these redshifts, ...

%FLARES (Wilkins et al. 2023), - removed because these seem to be stellar metallicities, not gas-phase?
In the bottom panel of Figure \ref{fig:MZR}, we also compare the observed MZR with predictions from cosmological simulations at $z\sim7$, including from \texttt{NEWHORIZON} (\citealt{dubois_introducing_2021}), TNG100 (\citealt{torrey_evolution_2019}), First Light (\citealt{nakazato_simulations_2023}), FIRE-2 (\citealt{marszewski_high-redshift_2024}), SPHINX (\citealt{katz_sphinx_2023}), and SERRA (\citealt{pallottini_mass-metallicity_2024}). We also plot for comparison the predictions from the semi-analytical model, DELPHI (Mauerhofer et al. in prep, see also \citealt{Delphi}), which incorporates cold gas fractions and star-formation efficiencies derived from the SPHINX simulation.  Where necessary, we have scaled these relations to a \cite{kroupa_variation_2001} IMF and a solar metallicity of $12+\log(\mathrm{O/H})=8.69$ (\citealt{asplund_chemical_2009}). Across these different simulations, the slopes at $\log(M_*/\mathrm{M_\odot})\sim7.5$-$8.5$ are broadly consistent. Comparisons with these predictions at the low-mass end ($\log(M_*/\mathrm{M_\odot})\lesssim8$) are detailed in \cite{chemerynska_extreme_2024}, and here we focus on the high-mass end populated by the REBELS sources ($\log(M_*/\mathrm{M_\odot})\sim9$-$10$), where there are some variations in the predicted slope and normalisation of the MZR.

In particular, FirstLight and FIRE-2 predict similarly metal-rich, massive galaxies at these redshifts, consistent with the observations of the REBELS sample, although we note that both of these predictions show the redshift evolution of selected massive galaxies from zoom-in simulations, whilst for the other simulations the median and/or $1\sigma$ distribution of the MZR for the full sample within the large-volume of the simulations are reported at each analysed redshift in the corresponding papers, which typically contain few massive ($\log(M_*/\mathrm{M_\odot})\gtrsim9$) galaxies at $z>6$. In \cite{marszewski_high-redshift_2024}, from which we plot the best-fit MZR from the `Slope and Normalization Evolution' model at $z=7$,  34 massive galaxies from FIRE-2 (\citealt{ma_simulating_2018,ma_dust_2019}) are selected for zoom-in simulations run down to either $z=5, 7,$ or 9, with halo masses $\sim 10^9$-$10^{12}\mathrm{M_{\odot}}$ at these final redshifts. In \cite{nakazato_simulations_2023}, from which we plot the 5-95\% dispersion in the sample at $z=7$, 62 galaxies are selected from the FirstLight simulation suite (\citealt{ceverino_introducing_2017}) with maximum circular velocities $>178$ km s$^{-1}$ at $z=5$. Differences in feedback prescriptions and stellar yields may also contribute to variations, and we add there may be other systematic differences in the way each simulation derives the stellar masses and metallicities. For example, the mass-weighted metallicity is given for FIRE-2 and \texttt{NEWHORIZON}. However, in FirstLight, SPHINX and TNG100, the gas-phase metallicities are weighted by either [O \textsc{iii}] line luminosity or SFR. This likely makes their predictions more comparable to observations, which are only sensitive to the bright, line-emitting star forming regions. 

A full assessment of the systematic differences between different simulations is beyond the scope of this work, but overall the consistency with simulations like FirstLight and FIRE-2 supports the idea that massive galaxies in the reionization era could reach relatively high metallicities in a short timescale. These findings align with a scenario where substantial chemical evolution occurs early and rapidly in massive galaxies, within just a few hundred million years.

\subsection{The Fundamental Metallicity Relation (FMR)}
\label{sec:fmr}

\begin{figure*}
    \centering
        \includegraphics[width=0.9\textwidth]{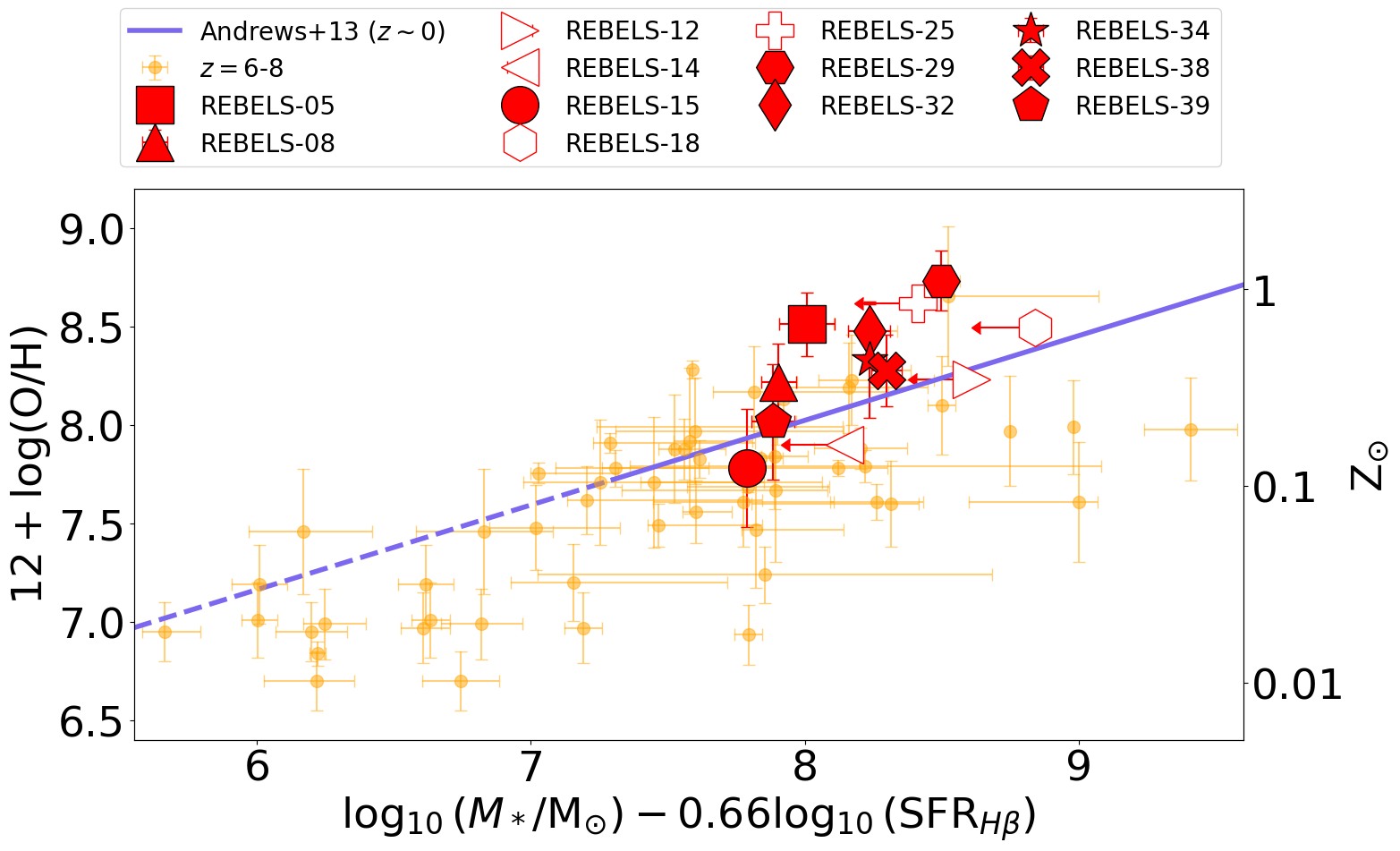}
    
    \caption{We plot the fundamental metallicity relation (FMR) for the REBELS galaxies and the literature sample at $z=6$-$8$ (\protect\citealt{nakajima_jwst_2023,chemerynska_extreme_2024}). We compare to the $z\sim0$ FMR from \protect\cite{andrews_mass-metallicity_2013} (solid blue line, extrapolated to $\mu_0.66<7.3$ with the dashed line). White-filled markers indicate the REBELS galaxies where SFR$_{\mathrm{H}\beta}$ is not attenuation-corrected, and is therefore a lower-limit.}
    \label{fig:FMR}
   
    \label{fig:fmr figures}
\end{figure*}

To address the scatter in the $z = 0$ MZR, the SFR was introduced as an additional variable in what is known as the FMR (e.g. \citealt{ellison_clues_2008,mannucci_fundamental_2010,lara-lopez_fundamental_2010}), where $12+\log(\mathrm{O/H})$ decreases with increasing SFR at fixed $M_*$. As with, for example, \cite{nakajima_jwst_2023}, we use the following to define the FMR:
\begin{equation}
\mu_{\alpha}=\log(M_*/\mathrm{M_{\odot}}) -\alpha \log(\mathrm{SFR/M_{\odot} yr^{-1}}),,
\end{equation}

\noindent where $\alpha = 0.66$ is found to minimise the scatter in the FMR at $z\sim0$ in \cite{andrews_mass-metallicity_2013} (solid blue line in Figure \ref{fig:FMR}). 

The FMR has been found to be largely redshift invariant out to $z\sim4$ (e.g. \citealt{henry_low_2013, sanders_mosdef_2021,curti_mass-metallicity_2020,henry_mass-metallicity_2021,nakajima_jwst_2023,heintz_dilution_2023,curti_chemical_2023,langeroodi_evolution_2023}). At $z>4$, however, recent studies have found a deviation in the observed FMR, although the exact redshift at which the observations deviate from the $z\sim0$ FMR differs across the literature (e.g., $z\sim7$ in \citealt{heintz_dilution_2023}, $z\sim8$ in \citealt{nakajima_jwst_2023}, $z\sim6$ in \citealt{curti_jades_2023}, $z\sim4$ in \citealt{langeroodi_evolution_2023}). These studies find that high-$z$ galaxies are more metal-poor than the $z\sim0$ FMR predicts, however with a poor sampling of high-mass galaxies out to higher redshifts. With the REBELS galaxies observed here, we extend this study to more massive EoR galaxies.

In Figure \ref{fig:FMR}, we find that the REBELS galaxies studied in this work typically exhibit higher values of $\mu_{0.66}$, which corresponds to lower sSFR compared to other $6<z<8$ galaxies in the literature. These REBELS sources are also scattered above and below the $z\sim0$ FMR, whereas most of the literature sample lies below it. This could suggest that the predominantly negative FMR offsets reported for high-$z$ galaxies may partially reflect a selection bias in current studies, which often target low-mass, metal-poor galaxies with high sSFRs and [O \textsc{iii}] EWs that may be more likely to show negative residuals relative to the local FMR.

Extreme line emitters, characterised by large [O \textsc{iii}] EWs and high sSFRs, may be more likely to show negative residuals relative to the FMR because their ISM conditions deviate significantly from the typical balance of star formation, metallicity, and gas content seen in the local Universe. These systems likely experience recent bursts of star formation that dominate the observed spectra, amplifying the appearance of low metallicity and high ionisation regions.

Another plausible explanation for a negative offset to the FMR at high-$z$ is significant accretion of pristine gas from the IGM at these redshifts, which dilutes the chemical abundances in the ISM and causes galaxies to be more metal-poor than the FMR predicts (\citealt{heintz_dilution_2023}). In this scenario, the REBELS galaxies could represent a more evolved population that, despite their high gas masses and fractions ($\log M_{\mathrm{gas}}/\mathrm{M_{\odot}}=9.7$-$10.7$, $f_{\mathrm{gas}}=0.73$-$0.96$, Algera et al. submitted) indicative of ongoing accretion, earlier bursts of star formation may have already efficiently built up their stellar mass and enriched their ISM. Currently, only galaxies in the REBELS sample and a handful of other massive $z>6$ galaxies have independent measurements of both gas mass ([C\textsc{ii}]-based) and metallicity (e.g. RXCJ0600-z6.3 at $z = 6.02$ from \citealt{fujimoto_primordial_2024} and \citealt{valentino_cold_2024}; S04950 at $z = 8.50$ from \citealt{heintz_gas_2023} and \citealt{fujimoto_jwst_2024}; and MACS1149-JD1 at $z = 9.11$ from \citealt{laporte_absence_2019} and \citealt{morishita_accelerated_2024}), significantly limiting investigations into the role of gas accretion in shaping the FMR at these redshifts. To distinguish between these possibilities, a larger sample of high-$z$ galaxies with robust measurements of gas mass, metallicity, stellar mass, and star formation rate -- particularly more gas mass estimates for low-mass systems -- would be essential.

We also caution that there are various systematics in determining these properties, and we note that the \cite{andrews_mass-metallicity_2013} $z\sim0$ FMR is derived from direct method metallicities, although similar conclusions can be drawn when using the FMR from metallicities derived using strong-line methods in \cite{sanders_mosdef_2021}. In addition to the caveats already discussed for deriving the oxygen abundance (Section \ref{sec:caveats}), we were also unable to determine a nebular attenuation for four of the REBELS galaxies (white markers), which means that current estimates of the SFR from the uncorrected H$\beta$ luminosity can only be taken as lower limits, whilst it is also possible that their stellar masses could be underestimated due to degeneracies between stellar mass, age, and interstellar reddening. %Furthermore, all the stellar masses in the sample (from both REBELS and the literature) were obtained from integrated SED fitting. Recent studies with spatially resolved SED fitting have shown that integrated stellar masses may be underestimated due to the `outshining' effect (e.g. \citealt{gimenez-arteaga_spatially_2023}). Spatially resolved SED fitting of the REBELS galaxies will be the focus of subsequent works.

\section{Summary \& Conclusions}
\label{sec:conclusions}

In this work, we have investigated the emission line diagnostics and ionised gas properties of a sample of 12 massive EoR galaxies at $z=6.496$ to $7.675$ selected from the REBELS ALMA large program, observed with the \textit{JWST} NIRSpec/IFU in the prism mode. Our findings reveal that the REBELS galaxies exhibit low O2, O32, and Ne3O2 ratios compared to other high-$z$ galaxies from pre-existing large surveys, suggesting distinct ionising conditions. These ratios imply that the REBELS galaxies have less intense ionising fields than galaxies in existing spectroscopic surveys at these redshifts. Recent \textit{JWST} observations frequently identify galaxies with high O32 ratios (indicative of low metallicities), as these samples are often biased toward sources with extreme [O \textsc{iii}] EWs. By contrast, the REBELS sources have [O \textsc{iii}] EWs more comparable to the median of 450 \AA ~at $z\sim7$ (\citealt{labbe_spectral_2013}), enabling a more representative sampling of the ISM conditions of the general galaxy population at high redshift.

To derive the metallicities of our sample, we investigated a variety of strong-line indices and calibrations, and adopt the \citet{sanders_direct_2024} calibrations based on 46 auroral line detections at $z>2$ for our fiducial estimates. We predominantly used the R23 index, but used R3 where the fluxes could not be attenuation-corrected (for four galaxies where H$\alpha$ is redshifted out of the NIRSpec coverage), and used O32 where the R23 index provides no real solution (for three galaxies that show elevated R3 and R23 ratios). These calibrations reveal a median metallicity of $\sim0.4 \mathrm{Z_\odot}$, with five REBELS galaxies having $Z_{\mathrm{gas}}\gtrsim0.5\mathrm{Z_\odot}$. Using O2, O32, and Ne3O2 ratios to derive the oxygen abundance tends to result in lower oxygen abundances for this sample, which may be due to the primary dependence of these ratios on other properties such as the ionisation parameter, electron density, and sSFR. However, abundances derived using these indices still result in the REBELS galaxies being more metal-rich than the bulk of the $z>6$ population in the literature (but see e.g. \citealt{shapley_aurora_2024}). The discrepancies found across different calibrations call for a further study of metallicity calibrations at high-$z$, and in particular calls for an increase in auroral line detections at higher metallicities.

The existence of metal-rich galaxies in the early Universe provides key insights into the rapid evolution of galaxies during the EoR. The REBELS sample suggests that some high-mass galaxies at $z \gtrsim 6$ exhibit relatively high metallicities, indicating that metal enrichment can rapidly occur within a few hundred million years of the Big Bang. Supporting this notion, a handful other galaxies from the reionisation era have also been found to display similarly low O32 ratios (\citealt{killi_solar_2022, witten_rising_2024,shapley_aurora_2024}), and several recent studies have identified evolved stellar populations at $z \gtrsim 6$ (e.g., \citealt{kuruvanthodi_strong_2024}). In addition, results from the FirstLight (\citealt{nakazato_simulations_2023}) and FIRE-2 (\citealt{marszewski_high-redshift_2024}) simulations indicate that some massive galaxies can be enriched to near-solar abundances at these early cosmic times. These findings hint at a potentially non-negligible population of evolved galaxies within the EoR, suggesting that some galaxies may have experienced significant metal enrichment and advanced stages of stellar evolution much earlier than previously expected.

By compiling literature results from lower-mass samples at the same redshift ($z=6$–$8$), we analysed the MZR and FMR over a $\sim4$ dex range in stellar mass. Notably, the REBELS galaxies significantly improve the sampling of the high-mass end of the MZR, more than doubling the number of massive ($\log(M_*/\mathrm{M_\odot}) > 9$) star-forming galaxies with metallicity estimates at $z>6$. The best-fit MZR derived, with a slope of $\gamma = 0.37\pm0.03$, is consistent with some previous studies at $z>6$ (e.g., \citealt{heintz_dilution_2023,chemerynska_extreme_2024}) and is steeper than some studies at lower redshifts (e.g., \citealt{sanders_mosdef_2021}). However, we note that systematic differences in the derivation of stellar masses, oxygen abundances, and SFRs across the literature may impact these comparisons.  

For the FMR, the REBELS galaxies show a mix of positive and negative offsets relative to the $z=0$ relation, in contrast to the predominantly negative offsets reported for lower-mass, metal-poor galaxies in other high-$z$ studies. This distinction may arise from a selection bias in current $z>6$ samples, which often target extreme line emitters. These negative offsets may reflect significant accretion of pristine gas from the IGM, diluting ISM metallicities and driving galaxies below the $z=0$ FMR. In this context, the REBELS galaxies could represent a more evolved population that has already undergone substantial enrichment through early efficient star formation while maintaining high gas fractions indicative of ongoing accretion. These findings highlight the need for a larger, more diverse sample of $z=6$–$8$ galaxies with robust measurements of stellar mass, gas mass, SFR, and metallicity to disentangle the effects of sample selection, ISM conditions, and galaxy evolution on the observed MZR and FMR at these redshifts.

\section*{Acknowledgements}
The authors would like to thank Alice Shapley for extensive discussions on high-$z$ metallicity calibrations. The authors would also like to thank Elisa Cataldi for discussions on metallicity calibrations with JWST data. LR acknowledges a grant from the Leiden University Fund/Bouwens Astrophysics Fund, www.luf.nl. MA acknowledges support from ANID Basal Project FB210003 and ANID MILENIO NCN2024\_112. HA and HI acknowledge support from the NAOJ ALMA Scientific Research Grant Code 2021-19A. AF acknowledges support from the ERC Advanced Grant INTERSTELLAR H2020/740120. JH acknowledges support from the ERC Consolidator Grant 101088676 (VOYAJ). PD warmly thanks the European Commission's and University of Groningen's CO-FUND Rosalind Franklin program.

%%%%%%%%%%%%%%%%%%%%%%%%%%%%%%%%%%%%%%%%%%%%%%%%%%
\section*{Data Availability}
The data used in this manuscript will be made available upon reasonable request to the corresponding author.



\bibliographystyle{mnras}

\bibliography{MZR_5}

%%%%%%%%%%%%%%%%%%%%%%%%%%%%%%%%%%%%%%%%%%%%%%%%%%

%%%%%%%%%%%%%%%%% APPENDICES %%%%%%%%%%%%%%%%%%%%%

\appendix

\section{Emission line fitting}
\label{sec:appendix emission line fitting}

When fitting the emission lines, we fix the emission line widths using the LSF derived in Stefanon et al. (in prep) and the line centroids according to the [C \textsc{ii}] redshift, accounting for a small wavelength offset in the NIRSpec observations, as detailed in Stefanon et al. in prep. We show the emission line-fitted, continuum-subtracted integrated spectrum and residuals for each galaxy in Figures \ref{fig:sub1} and \ref{fig:sub2}.

\begin{figure*}
  \begin{minipage}{\textwidth}
    \centering
    \includegraphics[width=.48\textwidth]{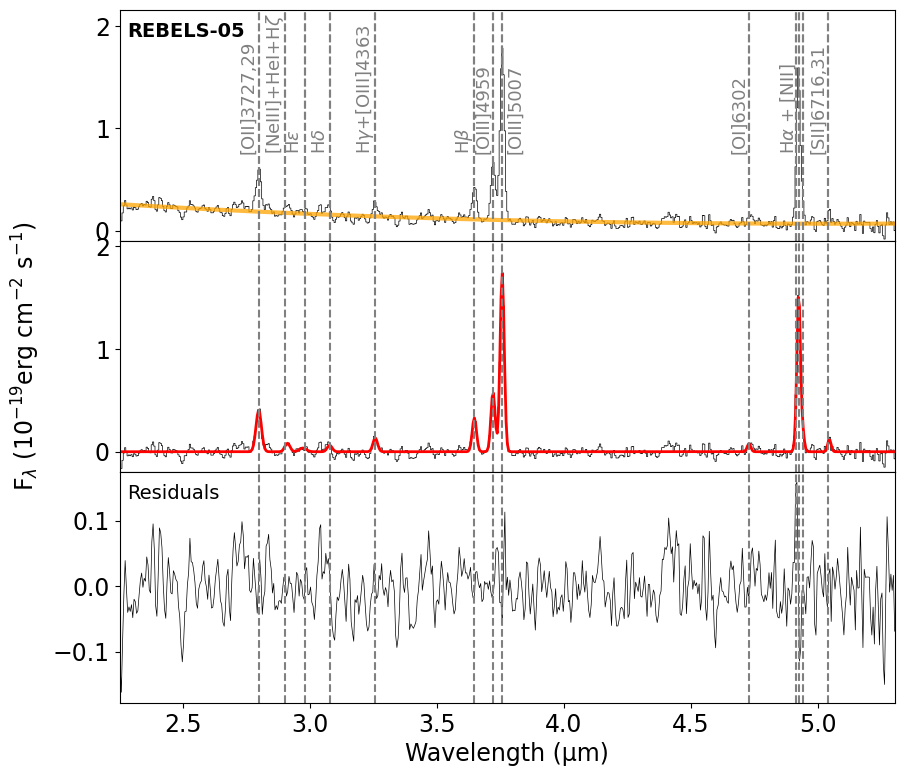}\quad
    \includegraphics[width=.48\textwidth]{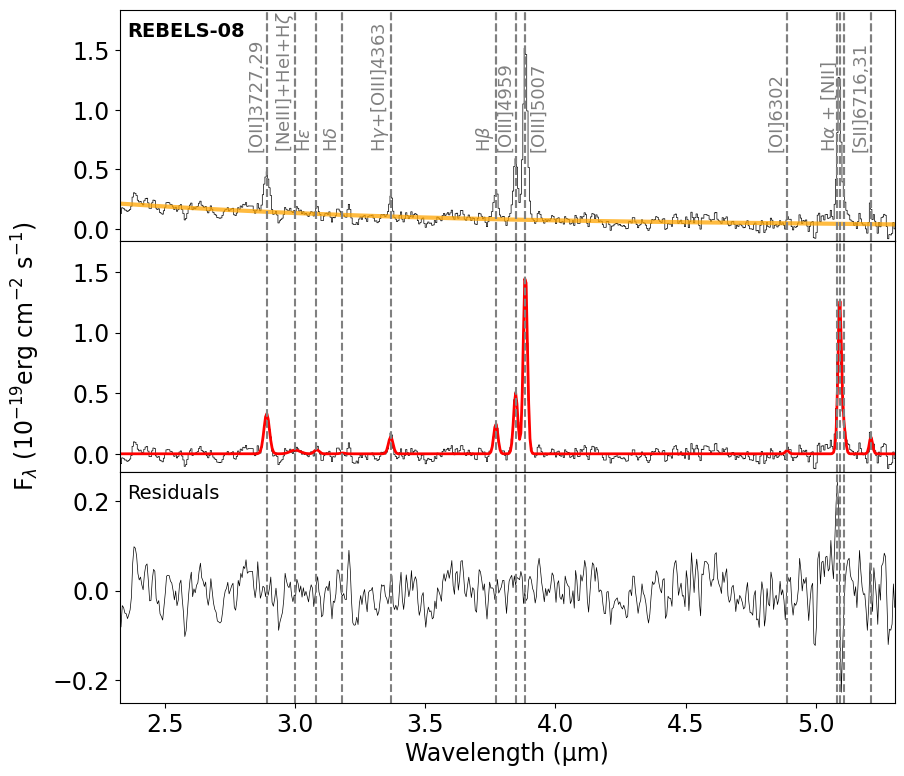}\\
    \includegraphics[width=.48\textwidth]{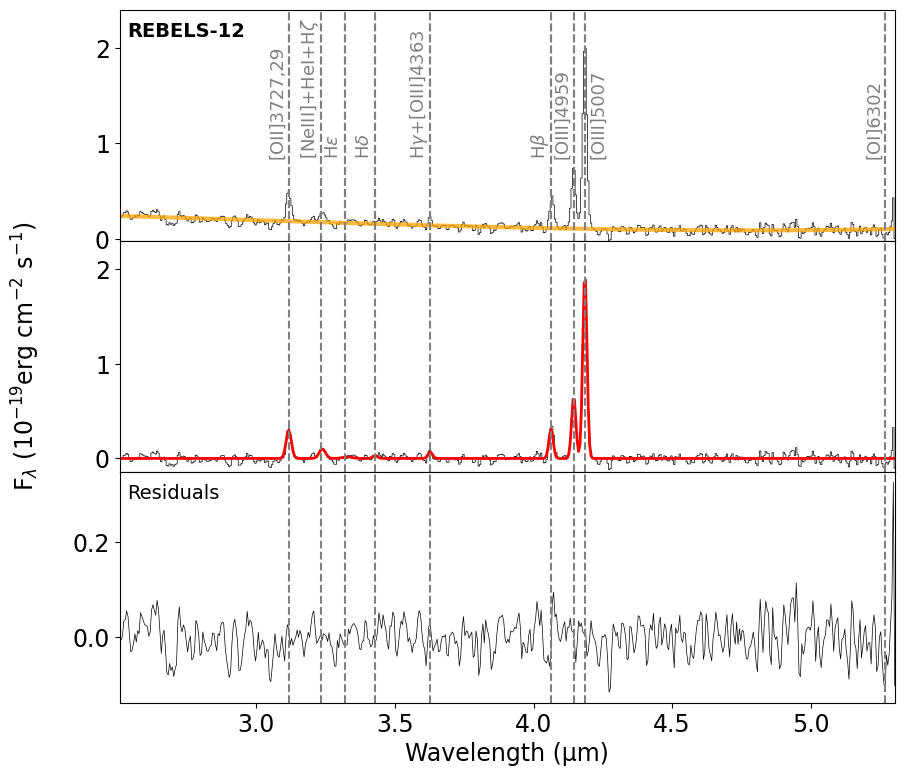}\quad
    \includegraphics[width=.48\textwidth]{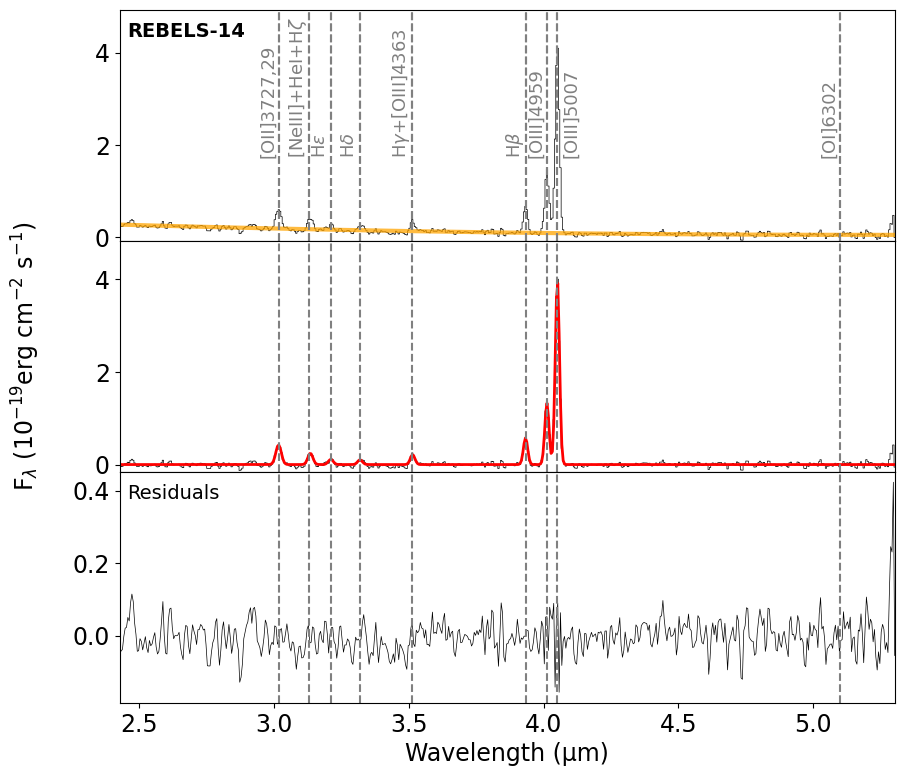}
    \includegraphics[width=.48\textwidth]{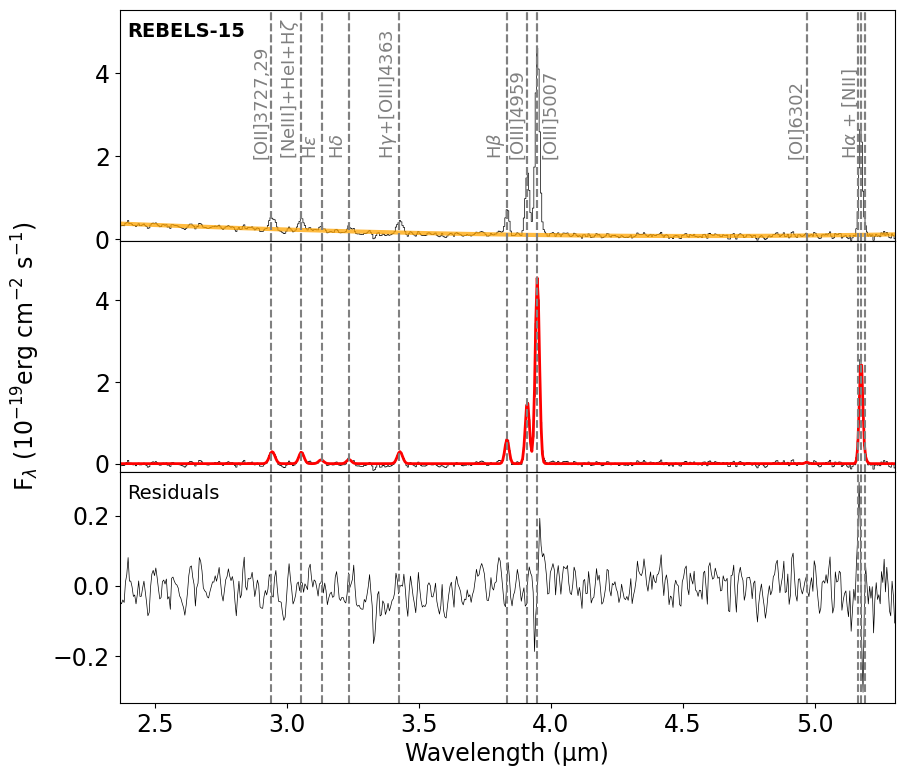}\quad
    \includegraphics[width=.48\textwidth]{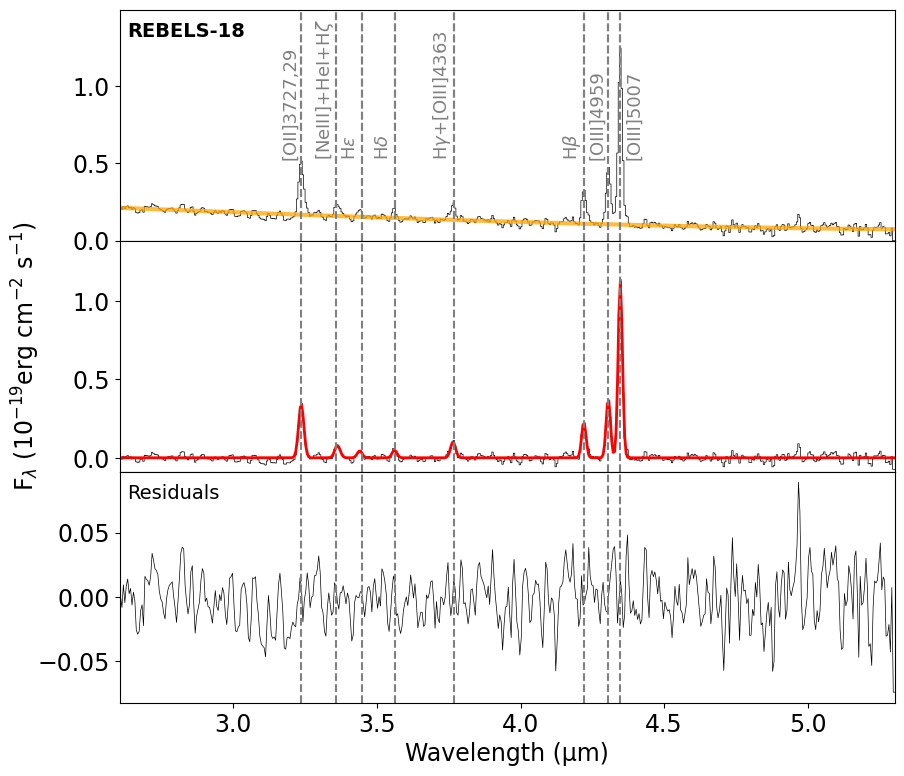}
    \caption{For each subfigure, we show a portion (rest-frame wavelength $>0.3\mu$m) of the observed spectrum of each galaxy (labelled in the top-left corner) which covers the key rest-frame optical nebular emission lines studied in this paper. For each galaxy, we plot the observed spectrum in grey and the \texttt{PYSPECKIT} continuum fit in orange in the top panel. In the middle panel, the Gaussian fits to each emission line in the continuum-subtracted spectrum are plotted in red, and in the bottom panel we plot the residuals of our emission line fitting.}
    \label{fig:sub1}
  \end{minipage}\\[1em]
\end{figure*}

\begin{figure*}
  \begin{minipage}{\textwidth}
    \centering
    \includegraphics[width=.48\textwidth]{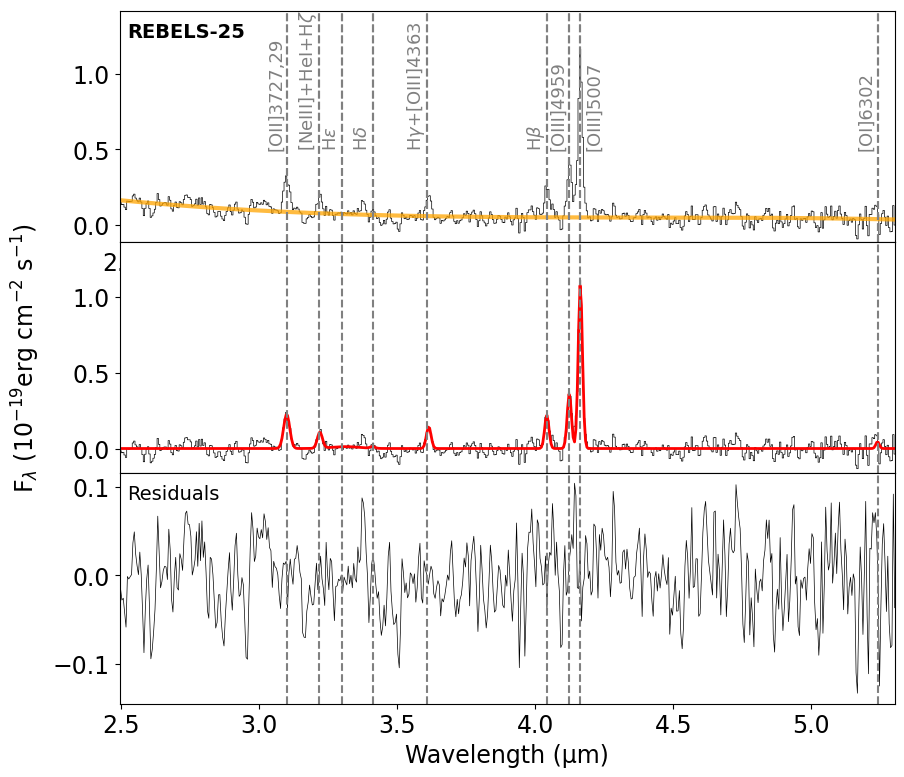}\quad
    \includegraphics[width=.48\textwidth]{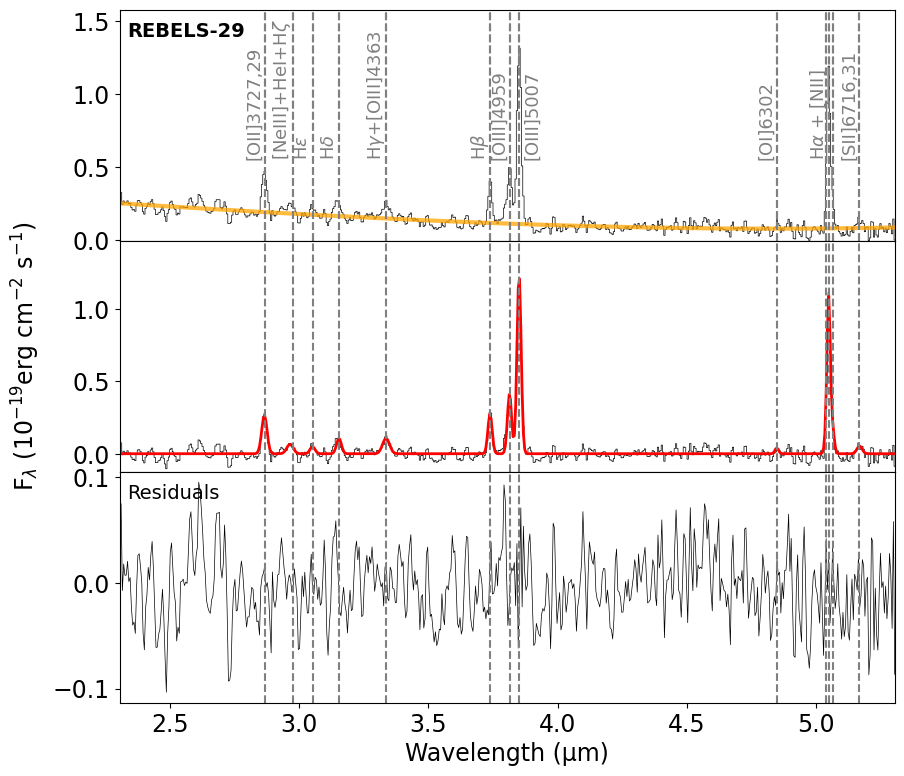}\\
    \includegraphics[width=.48\textwidth]{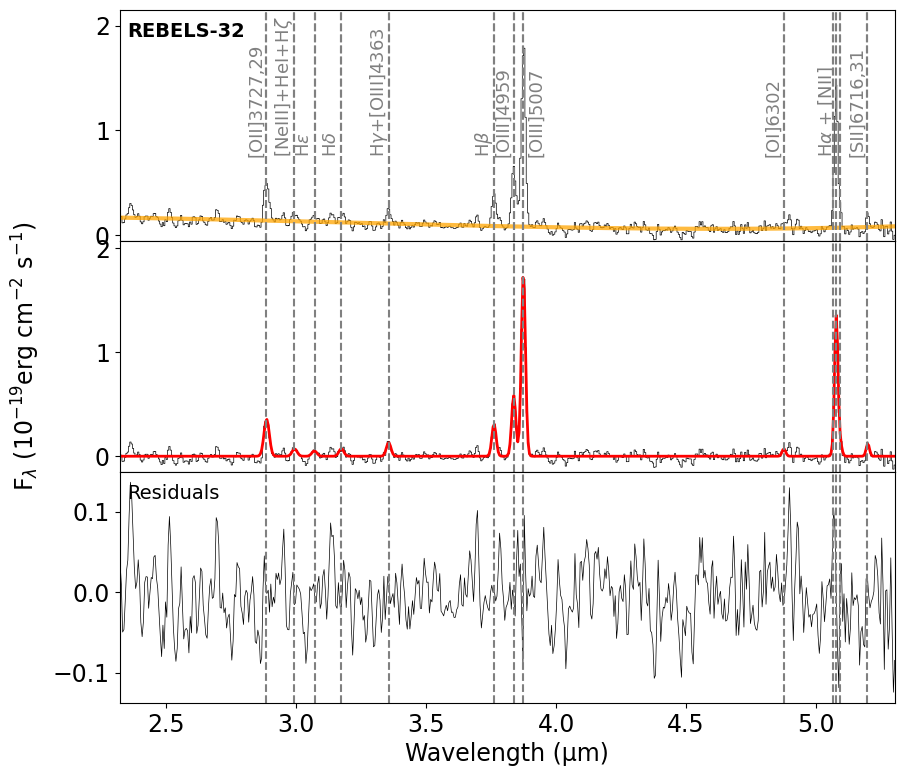}\quad
    \includegraphics[width=.48\textwidth]{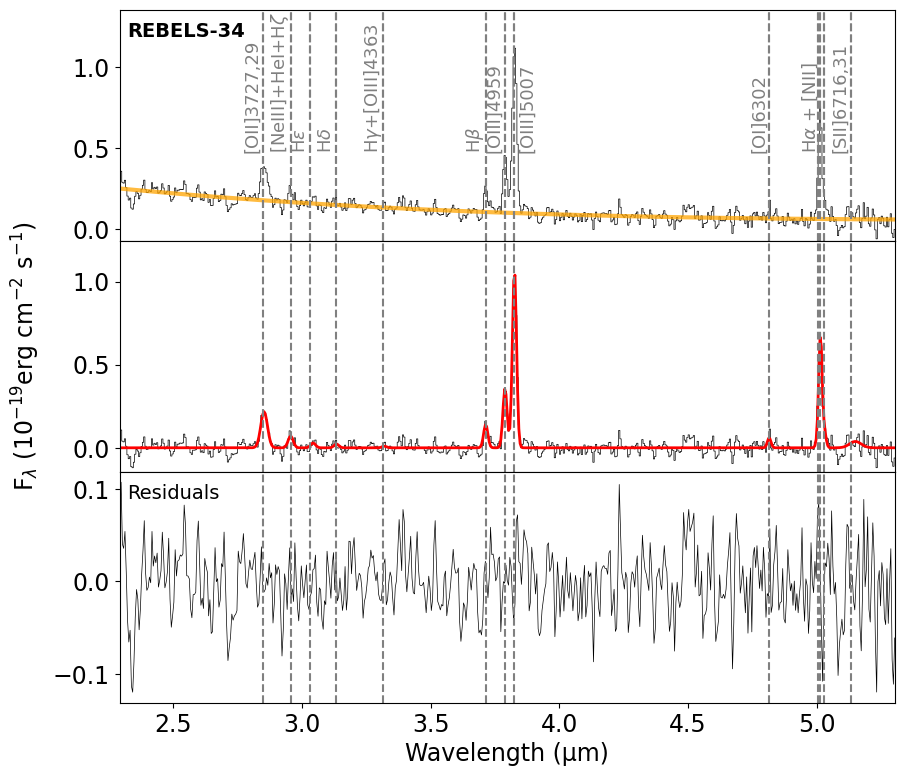}
    \includegraphics[width=.48\textwidth]{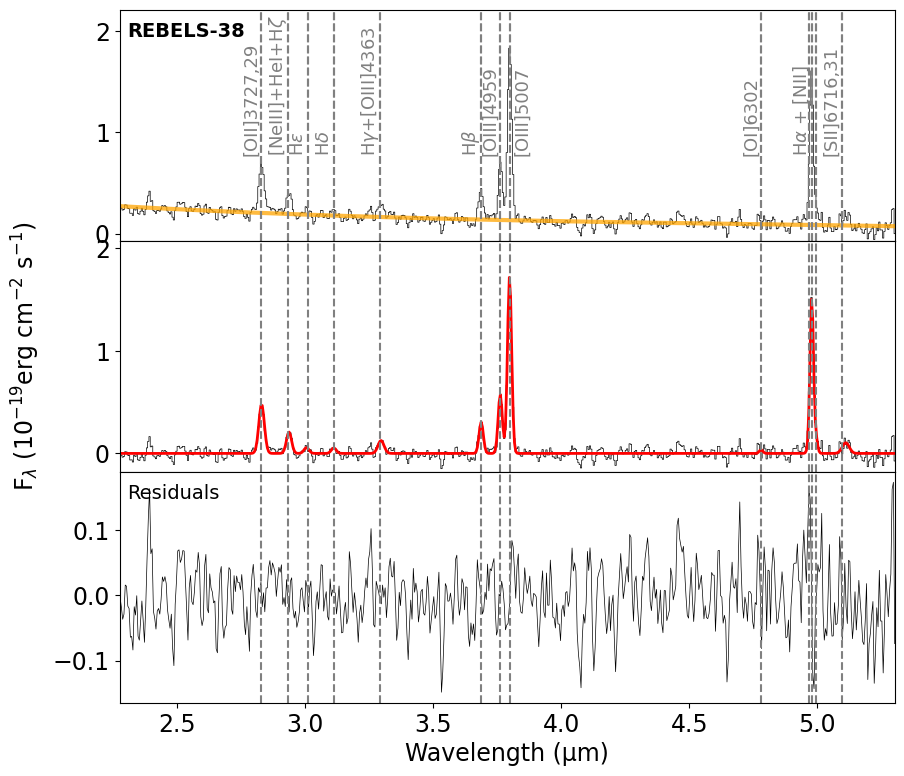}\quad
    \includegraphics[width=.48\textwidth]{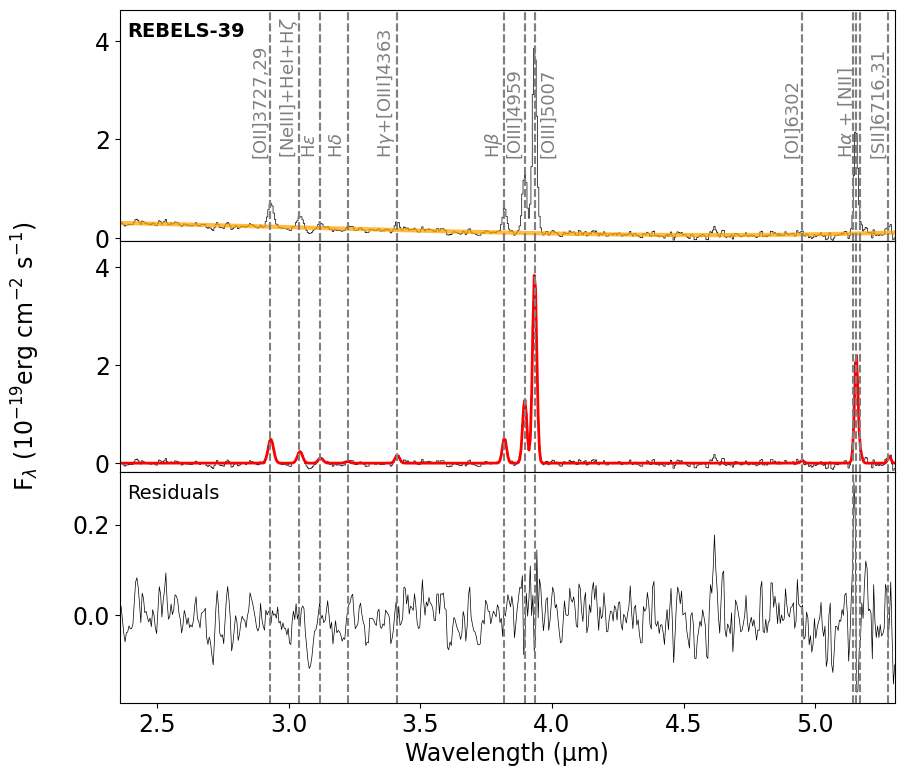}
    \caption{As with \ref{fig:sub1}.}
    \label{fig:sub2}
  \end{minipage}\\[1em]
\end{figure*}

\section{Balmer decrements}
\label{sec:appendix balmer decrements}

For 8/12 of the REBELS galaxies, we have the wavelength coverage to detect H$\alpha$ emission. For six of these, we detect H$\gamma +$[O \textsc{iii}]$\lambda$4363 emission at $>3\sigma$. Assuming that the contribution of the auroral [O \textsc{iii}]$\lambda$4363 line to this emission is negligible in comparison to the uncertainties, we compare the $A_{\mathrm{V, ~neb}}$ determined from the H$\alpha$/H$\beta$ ratio to that from the H$\gamma$/H$\beta$ ratio, using an intrinsic ratio of 0.47 for H$\gamma$/H$\beta$ (also following Case B recombination, \citealt{osterbrock_astrophysics_2006}). The derived $A_{\mathrm{V, ~neb}}$ are consistent within the uncertainties for four of these six galaxies, however we find that the H$\gamma$/H$\beta$ ratios are consistent with unphysical values (assuming Case B recombination) within the 1$\sigma$ uncertainties for all six (Figure \ref{fig:Balmer decrement}). This could be due to a non-negligible contribution from [O \textsc{iii}]$\lambda$4363 emission (and/or [Fe \textsc{ii}]$\lambda4360$, which dominates over the auroral line in metal-rich galaxies, \citealt{curti_new_2017, shapley_aurora_2024}), causing the H$\gamma$/H$\beta$ to be greater than 0.47, or due to the low SNR of these detections. Similarly, other studies have also found sources which appear to deviate from Case B recombination (e.g. \citealt{pirzkal_next_2024, scarlata_universal_2024}). Due to these uncertainties with using the H$\gamma$/H$\beta$ ratio to determine the attenuation, we do not attempt a reddening correction of the emission line fluxes based on the Balmer decrement for the four galaxies at $z\gtrsim7$, where H$\alpha$ is beyond the wavelength coverage of NIRSpec. We therefore give their non-attenuation-corrected values in Table \ref{tab:non-corrected flux catalogue}.

\begin{figure}[h]
    \centering
    \includegraphics[width=0.48\textwidth]{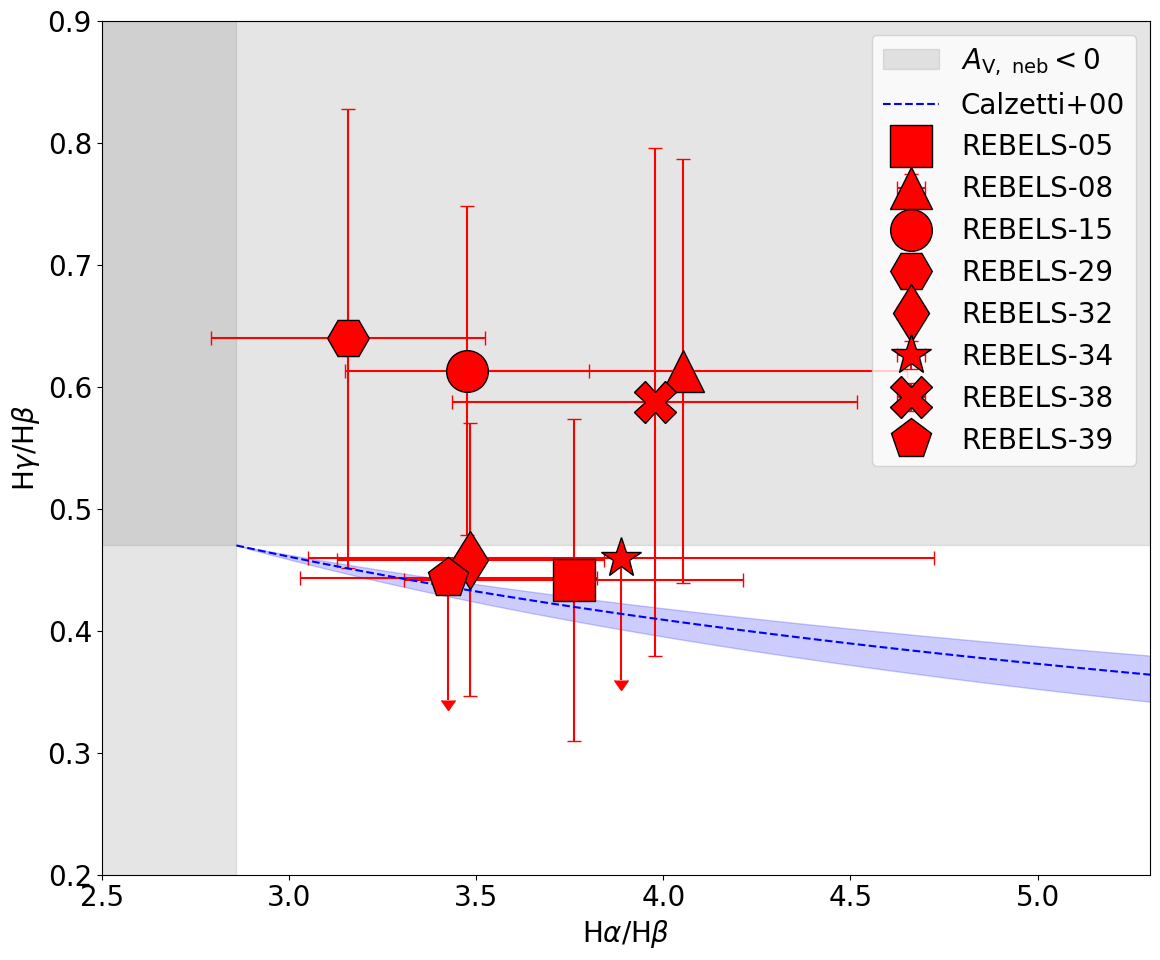}
    \caption{For 8/12 targets, we have the wavelength coverage for H$\alpha$, H$\beta$, and H$\gamma$. We plot a comparison of the H$\gamma$/H$\beta$ and H$\alpha$/H$\beta$ Balmer decrements for these eight sources, and compare to the expected trend assuming \protect\cite{calzetti_dust_2000} dust attenuation with theoretical values of H$\gamma$/H$\beta=0.47$ and H$\alpha$/H$\beta=2.86$. The H$\gamma$/H$\beta$ ratios of these REBELS galaxies are consistent with unphysical values (assuming Case B recombination) of the nebular attenuation (shown with the grey shading), likely due to the low SNR of H$\gamma$. We therefore do not use the H$\gamma$/H$\beta$ to attenuation-correct the line fluxes of the four sources where H$\alpha$ falls outside of our wavelength coverage, and we instead report their attenuation-uncorrected fluxes in Table \ref{tab:non-corrected flux catalogue}. }
    \label{fig:Balmer decrement}
\end{figure}

\section{Metallicity estimates with different calibrations}
\label{sec:appendix_metallicity}
\begin{figure}
    \centering
    \includegraphics[width=0.48\textwidth]{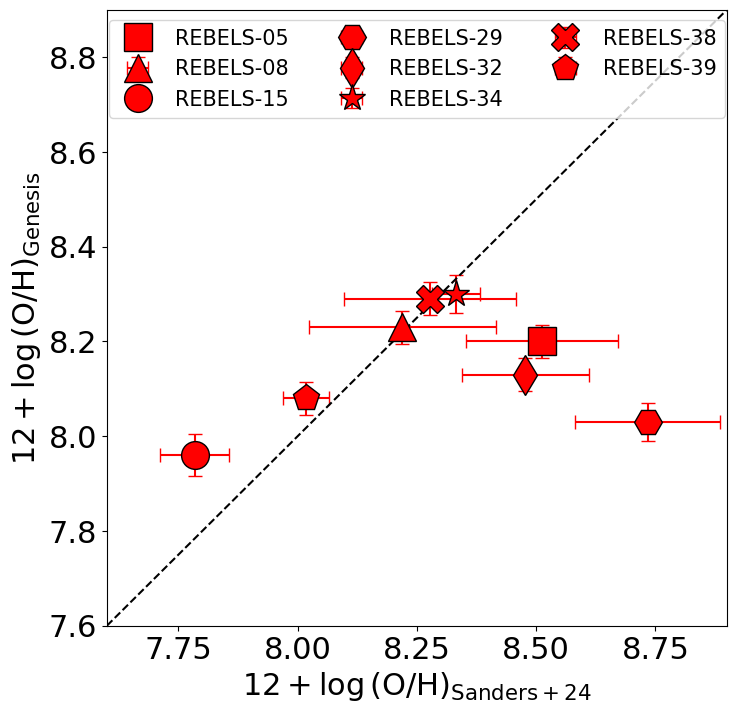}
    \caption{A comparison between the fiducial metallicity estimates using the empirically-derived calibrations in \protect\cite{sanders_direct_2024} with metallicity estimates using the non-parametric \texttt{genesis} method (\protect\citealt{langeroodi_genesis-metallicity_2024}).}
    \label{fig:genesis}
\end{figure}
Due to the considerable discrepancies found between different strong line indices and calibrations used to derive oxygen abundances in the literature, we re-derive the oxygen abundances of the REBELS sample using a variety of different calibrations, which we list in Tables \ref{tab:full sanders} to \ref{tab:bian_table}. For the \cite{nakajima_empress_2022} calibrations, we show results from both the high equivalent width (EW$>200$\AA) calibrations (calibrated for $12+\log(\mathrm{O/H})\sim 7.1$-$8.1$), and the calibrations based on the sample including stacked SDSS galaxies at higher metallicities (calibrated for $12+\log(\mathrm{O/H})\sim 7.0$-$8.9$) collected therein. Although some of our H$\beta$ EWs are $<100$\AA, our observed R23, R3 and O32 ratios are above the cut-off of the small-EW calibration range, and the medium EW calibrations are not well-behaved due to the small sample size. Similarly, in \cite{laseter_jades_2024}, the sample show higher ionisation and excitation ratios at a given metallicity than local galaxies with rest-frame H$\beta$ EW$=200$-$300$\AA, but have $\sim 2\times$ lower EWs, implying that H$\beta$ EW may not be a good probe of excitation conditions at this redshift, when making comparisons to lower redshifts.

We also used the calibration derived in \cite{laseter_jades_2024}, which uses $\hat{R}=0.47\log_{10} R2 + 0.88 \log_{10} R3$. However, we obtain real solutions for only two galaxies with attenuation-corrected fluxes in the REBELS sample, which are for REBELS-15 and REBELS-29. For REBELS-15, the two solutions with the $\hat{R}$ calibration are 7.95 and 8.27, and for REBELS-29 they are 7.79 and 8.39.  

In addition, we use the non-parametric gas-phase metallicity estimator, \texttt{genesis} (\citealt{langeroodi_genesis-metallicity_2024}) for the 8/12 galaxies where H$\alpha$ is detected. \texttt{genesis} uses inputted R3, R2, and EW(H$\beta$) values to estimate the oxygen abundance, and is calibrated on a sample of 1551 [OIII]$\lambda4363$ detections at $0<z<10$, although $>75\%$ are at $z\lesssim 1$. This sample includes the galaxies in each of the calibration samples of the aforementioned metallicity calibration papers. We plot the comparison between the fiducial oxygen abundances derived from the \cite{sanders_direct_2024} calibrations with those estimated with \texttt{genesis} in Figure \ref{fig:genesis}. We find that these metallicity estimates very closely align with the \citealt{sanders_direct_2024} O32 and O2 estimates in Table \ref{tab:full sanders}. 

From this sample of galaxies, it is clear that there are significant offsets, by as much as $\sim0.6$ dex, between different commonly used strong-line metallicity calibrations. However, when re-analysing the MZR and FMR with R3, R23, and O32 indices across the various calibrations, we find that the choice of calibration has little effect on the slope, intercept, or scatter of these relations. For example, we show in Figure \ref{fig:o32 mzr} the MZR using only the galaxies with attenuation-corrected O32 ratios and using the O32 metallicity calibration from \cite{sanders_direct_2024}. By fitting the MZR according to Equation \ref{eq:MZR}, we find a slope $\gamma=0.27\pm0.07$ and an intercept $Z_{10}=8.17\pm0.08$, which is consistent within the uncertainties with the values reported in Section \ref{sec:mzr}.

\begin{table*}
\caption{Oxygen abundances derived from various strong line indices using the \protect\cite{sanders_direct_2024} calibrations. Missing values indicate either undetected emission lines, or no real solutions to the calibration.}
\begin{tabular}{lccccc}
\toprule
   Galaxy &              O3 &             R23 &           Ne3O2 &             O32 &              O2 \\
\midrule
REBELS-05 & $8.44 \pm 0.14$ & $8.51 \pm 0.16$ &               - & $8.29 \pm 0.29$ & $8.23 \pm 0.23$ \\
REBELS-08 & $8.29 \pm 0.17$ &  $8.22 \pm 0.22$ &               - & $8.31 \pm 0.29$ &  $8.30 \pm 0.23$ \\
REBELS-12 & $8.23 \pm 0.13$ &               - &    $8 \pm 0.28$ &               - &               - \\
REBELS-14 & $7.90 \pm 0.12$ &               - & $7.89 \pm 0.25$ &               - &               - \\
REBELS-15 &               - &               - & $7.67 \pm 0.27$ &  $7.78 \pm 0.30$ & $7.84 \pm 0.24$ \\
REBELS-18 &  $8.50 \pm 0.13$ &               - & $8.24 \pm 0.27$ &               - &               - \\
REBELS-25 & $8.62 \pm 0.17$ &               - & $8.02 \pm 0.27$ &               - &               - \\
REBELS-29 & $8.57 \pm 0.14$ & $8.73 \pm 0.15$ &               - & $8.18 \pm 0.29$ & $8.06 \pm 0.23$ \\
REBELS-32 & $8.38 \pm 0.13$ & $8.48 \pm 0.13$ &               - &  $8.20 \pm 0.29$ & $8.16 \pm 0.23$ \\
REBELS-34 &               - &               - &               - & $8.33 \pm 0.29$ & $8.43 \pm 0.24$ \\
REBELS-38 & $8.38 \pm 0.16$ & $8.28 \pm 0.18$ & $8.01 \pm 0.26$ & $8.39 \pm 0.29$ & $8.36 \pm 0.23$ \\
REBELS-39 &               - &               - & $7.95 \pm 0.26$ & $8.02 \pm 0.29$ & $8.09 \pm 0.23$ \\
\bottomrule
\end{tabular}
\label{tab:full sanders}
\end{table*}

\begin{table*}
\caption{Oxygen abundances derived from various strong line indices using the \protect\cite{nakajima_empress_2022} calibrations for the high EW, local analog sample, calibrated for $12+\log(\mathrm{O/H})\sim 7.1-8.1$. Missing values indicate either undetected emission lines, or no real solutions to the calibration.}
\begin{tabular}{lccccccc}
\toprule
   Galaxy &              O3 &             R23 &           Ne3O2 &             O32 &              O2 &              N2 &            O3N2 \\
\midrule
REBELS-05 & $8.54 \pm 0.19$ & $8.51 \pm 0.15$ &               - & $8.43 \pm 0.39$ & $8.38 \pm 0.37$ & $9.19 \pm 0.26$ & $9.07 \pm 0.49$ \\
REBELS-08 &   $8.40 \pm 0.2$ & $8.21 \pm 0.17$ &               - &  $8.44 \pm 0.40$ &  $8.45 \pm 0.40$ & $9.32 \pm 0.26$ &  $9.1 \pm 0.48$ \\
REBELS-12 & $8.35 \pm 0.18$ &               - & $8.18 \pm 0.42$ &               - &               - &               - &               - \\
REBELS-14 & $8.15 \pm 0.17$ &               - &  $8.08 \pm 0.40$ &               - &               - &               - &               - \\
REBELS-15 &               - &               - & $7.84 \pm 0.41$ &  $7.94 \pm 0.40$ & $7.99 \pm 0.43$ &               - &               - \\
REBELS-18 &  $8.60 \pm 0.18$ &               - & $8.35 \pm 0.42$ &               - &               - &               - &               - \\
REBELS-25 &  $8.71 \pm 0.20$ &               - &  $8.20 \pm 0.42$ &               - &               - &               - &               - \\
REBELS-29 & $8.66 \pm 0.19$ & $8.71 \pm 0.15$ &               - & $8.35 \pm 0.39$ & $8.22 \pm 0.36$ & $9.26 \pm 0.26$ & $9.14 \pm 0.48$ \\
REBELS-32 & $8.48 \pm 0.18$ & $8.48 \pm 0.14$ &               - & $8.36 \pm 0.39$ & $8.32 \pm 0.34$ & $8.95 \pm 0.28$ & $8.91 \pm 0.53$ \\
REBELS-34 &               - &               - &               - &  $8.46 \pm 0.40$ & $8.55 \pm 0.48$ &               - &               - \\
REBELS-38 & $8.48 \pm 0.19$ & $8.28 \pm 0.16$ &  $8.19 \pm 0.4$ &  $8.5 \pm 0.39$ &  $8.5 \pm 0.37$ & $9.08 \pm 0.28$ & $8.99 \pm 0.52$ \\
REBELS-39 &               - &               - & $8.14 \pm 0.41$ &  $8.21 \pm 0.40$ & $8.25 \pm 0.38$ &               - &               - \\
\bottomrule
\end{tabular}
\label{tab:high ew nakajima}

\end{table*}

\begin{table*}
\caption{Oxygen abundances derived from various strong line indices using the \protect\cite{nakajima_empress_2022} calibrations for the full sample, including stacked SDSS spectra at higher metalliities (range of $6.9<12+\log(\mathrm{O/H})<8.9$ for all but Ne3O2, which has range $7<12+\log(\mathrm{O/H})<8.1$). Missing values indicate either undetected emission lines, or no real solutions to the calibration.}
\begin{tabular}{lccccccc}
\toprule
   Galaxy &              R3 &             R23 &           Ne3O2 &             O32 &              O2 &              N2 &            O3N2 \\
\midrule
REBELS-05 & $8.15 \pm 0.17$ & $8.17 \pm 0.18$ &               - & $8.18 \pm 0.39$ & $8.26 \pm 0.27$ & $8.45 \pm 0.24$ & $9.03 \pm 0.42$ \\
REBELS-08 & $8.02 \pm 0.22$ &               - &               - & $8.19 \pm 0.39$ & $8.34 \pm 0.28$ &  $8.50 \pm 0.24$ & $9.02 \pm 0.42$ \\
REBELS-12 &               - &               - &               - &               - &               - &               - &               - \\
REBELS-14 &               - &               - & $8.93 \pm 0.58$ &               - &               - &               - &               - \\
REBELS-15 &               - &               - & $9.78 \pm 0.61$ &               - & $7.78 \pm 0.27$ &               - &               - \\
REBELS-18 & $8.19 \pm 0.17$ &               - &               - &               - &               - &               - &               - \\
REBELS-25 & $8.26 \pm 0.17$ &               - &               - &               - &               - &               - &               - \\
REBELS-29 & $8.23 \pm 0.17$ & $8.36 \pm 0.13$ &               - & $8.09 \pm 0.39$ & $8.07 \pm 0.27$ & $8.48 \pm 0.24$ & $9.01 \pm 0.42$ \\
REBELS-32 &  $8.10 \pm 0.17$ & $8.12 \pm 0.19$ &               - &  $8.10 \pm 0.39$ & $8.18 \pm 0.27$ & $8.36 \pm 0.24$ & $9.06 \pm 0.42$ \\
REBELS-34 &               - &               - &               - & $8.21 \pm 0.39$ &               - &               - &               - \\
REBELS-38 & $8.11 \pm 0.18$ &               - &               - & $8.25 \pm 0.39$ & $8.42 \pm 0.28$ & $8.41 \pm 0.24$ & $9.04 \pm 0.42$ \\
REBELS-39 &               - &               - &               - &  $7.91 \pm 0.40$ & $8.11 \pm 0.27$ &               - &               - \\
\bottomrule
\end{tabular}
\label{tab:full sample nakajima}

\end{table*}

\begin{table*}
\caption{Oxygen abundances derived from various strong line indices using the \protect\cite{bian_direct_2021} calibrations. Missing values indicate either undetected emission lines, or no real solutions to the calibration. We note that the reported uncertainties are only based on the propoagated uncertainties in the fluxes, since the uncertainty in the calibrations is not given in \protect\cite{bian_direct_2021}.}
\begin{tabular}{lcccccc}
\toprule
   Galaxy & R3 & R23 &            Ne3O2 &              O32 &               N2 &             O3N2 \\
\midrule
REBELS-05 &       - &   - &                - & $8.31 \pm 0.027$ & $8.42 \pm 0.049$ &  $8.3 \pm 0.049$ \\
REBELS-08 &       - &   - &                - &  $8.32 \pm 0.03$ & $8.49 \pm 0.043$ & $8.32 \pm 0.043$ \\
REBELS-12 &       - &   - & $8.04 \pm 0.085$ &                - &                - &                - \\
REBELS-14 &       - &   - & $7.97 \pm 0.041$ &                - &                - &                - \\
REBELS-15 &       - &   - & $7.84 \pm 0.071$ &  $7.97 \pm 0.05$ &                - &                - \\
REBELS-18 &       - &   - & $8.19 \pm 0.082$ &                - &                - &                - \\
REBELS-25 &       - &   - & $8.06 \pm 0.078$ &                - &                - &                - \\
REBELS-29 &       - &   - &                - & $8.24 \pm 0.027$ & $8.45 \pm 0.048$ & $8.34 \pm 0.048$ \\
REBELS-32 &       - &   - &                - &  $8.25 \pm 0.02$ & $8.29 \pm 0.069$ & $8.21 \pm 0.069$ \\
REBELS-34 &       - &   - &                - & $8.34 \pm 0.034$ &                - &                - \\
REBELS-38 &       - &   - & $8.05 \pm 0.054$ & $8.38 \pm 0.024$ & $8.36 \pm 0.064$ & $8.25 \pm 0.064$ \\
REBELS-39 &       - &   - & $8.01 \pm 0.069$ & $8.13 \pm 0.033$ &                - &                - \\
\bottomrule
\end{tabular}
\label{tab:bian_table}
\end{table*}

\begin{figure*}
    \centering
    \includegraphics[width=0.88\textwidth]{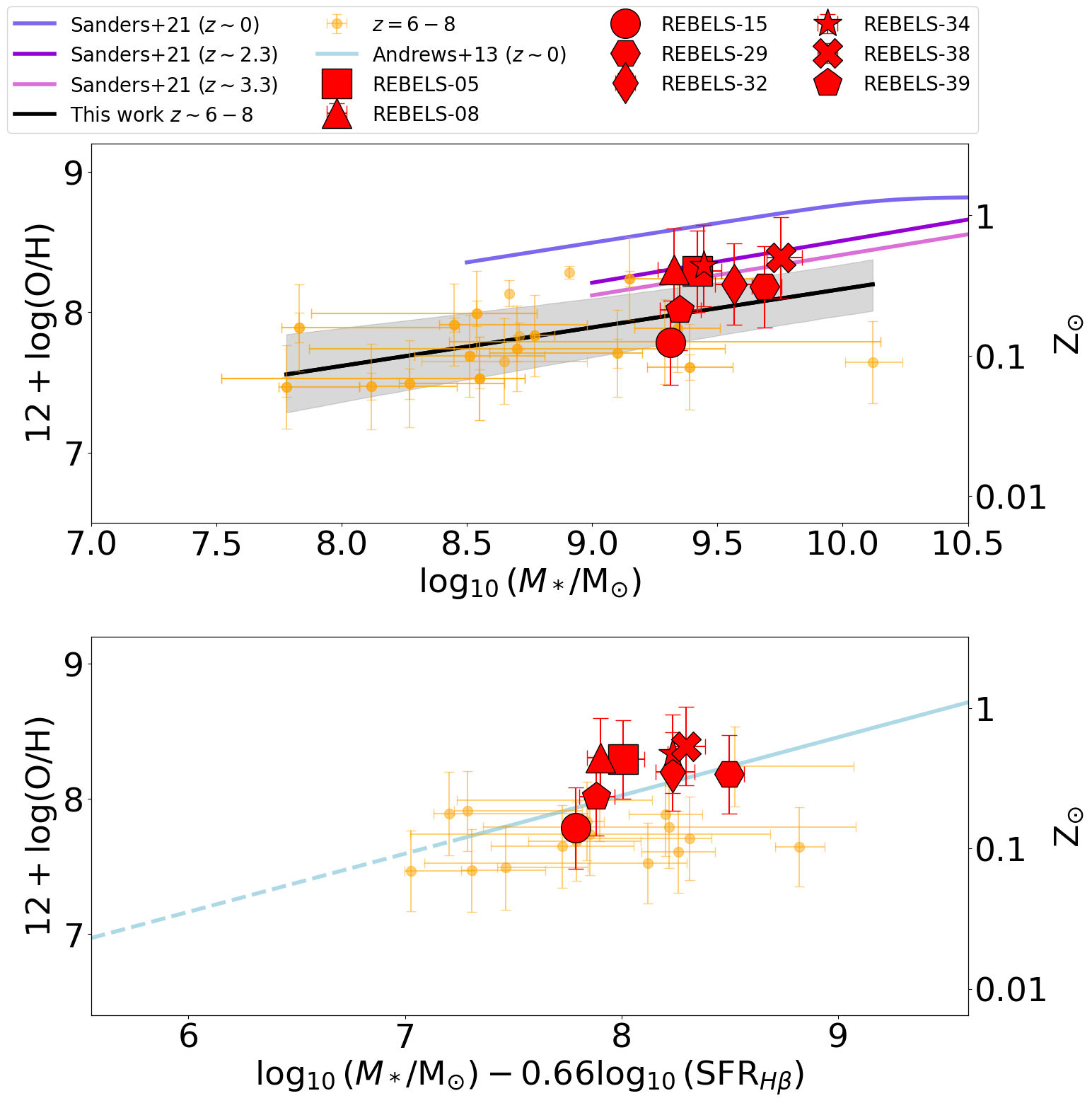}
    \caption{We plot the MZR (top panel) and FMR (lower panel) for the sources where the attenuation-corrected O32 ratio can be used to determine the metallicity. Markers, lines and shading are the same as in Figures \ref{fig:MZR} and \ref{fig:FMR}.}
    \label{fig:o32 mzr}
\end{figure*}

%%%%%%%%%%%%%%%%%%%%%%%%%%%%%%%%%%%%%%%%%%%%%%%%%%

% Don't change these lines
\bsp	% typesetting comment
\label{lastpage}
\end{document}